\documentclass[aps,prl,superscriptaddress,twocolumn,showpacs]{revtex4-1}

\usepackage[utf8]{inputenc}
\usepackage[T1]{fontenc}
\usepackage[german,english]{babel}

\usepackage{amsmath}
\usepackage{mathtools}
\usepackage{amssymb}
\usepackage{amsthm}
\usepackage{microtype} 
\usepackage{quoting} 
\quotingsetup{font=small}

\usepackage[pdftex]{graphicx}
\usepackage{fancyhdr}
\usepackage{bm}
\usepackage{comment}
\usepackage{booktabs}
\usepackage{eurosym} 
\usepackage{braket}
\usepackage{bbold} 
\usepackage{emptypage}
\usepackage{epstopdf}
\usepackage{verbatim}
\usepackage{enumitem}
\usepackage{float} 

\graphicspath{{Figures/PNG/}{Figures/PDF/}{Figures/EPS/}{Figures/TEX/}{Figures/}} 

\usepackage[usenames,dvipsnames]{xcolor} 
\definecolor{islamicgreen}{rgb}{0.0, 0.56, 0.0}

\newcommand{\beq}{\begin{equation}}
\newcommand{\eeq}{\end{equation}}
\newcommand {\nn} \nonumber

\renewcommand{\ket}[1]{\left|#1\right\rangle}
\renewcommand{\bra}[1]{\left\langle #1\right|}

\DeclarePairedDelimiter{\abs}{\lvert}{\rvert} 

\DeclareMathOperator{\Tr}{Tr}
\DeclareMathOperator{\Min}{Min}

\DeclareMathOperator{\arccoth}{arccoth}

\renewcommand{\sim}{\thicksim}

{\left\lbrace\begin{array}{@{}l@{}}}%
{\end{array}\right.}

\usepackage[bookmarks=true,colorlinks,linkcolor=blue,urlcolor=blue,citecolor=blue]{hyperref} 

\def\ro{\hat{\rho}}

\begin{document}

\title{
Origin of the slow growth of entanglement entropy \\ in long-range interacting {spin} systems} 

\author{Alessio Lerose} 
\affiliation{SISSA --- International School for Advanced Studies, via Bonomea 265, I-34136 Trieste, Italy}
\affiliation{INFN --- Istituto Nazionale di Fisica Nucleare, Sezione di Trieste, I-34136 Trieste, Italy}

\author{Silvia Pappalardi} 
\affiliation{SISSA --- International School for Advanced Studies, via Bonomea 265, I-34136 Trieste, Italy}
\affiliation{Abdus Salam ICTP --- International Center for Theoretical Physics, Strada Costiera 11, I-34151 Trieste, Italy}

\date{\today} 

\begin{abstract}
{Long-range interactions 
allow far-distance quantum correlations to build up very fast. 
Nevertheless, 
numerical simulations demonstrated a dramatic slowdown of entanglement entropy growth after a sudden quench.
In this work, we unveil the general mechanism underlying this counterintuitive phenomenon for 
$d$-dimensional quantum spin systems with slowly-decaying 
interactions.
We demonstrate that the semiclassical rate of collective spin squeezing governs the dynamics of entanglement, leading to a universal logarithmic growth in the absence of semiclassical chaos. 
In fact, the standard quasiparticle contribution is shown to get suppressed as the interaction range is sufficiently increased. 
All our analytical results agree with numerical computations for quantum Ising chains with long-range couplings. Our findings 
thus identify a qualitative change in the entanglement production induced by long-range interactions,
and 
are experimentally relevant for accessing entanglement in highly-controllable platforms, including trapped ions, atomic condensates and cavity-QED systems.}
\end{abstract}


\maketitle

\begin{figure}[H]
\centering
\vspace{-4.5mm}
\includegraphics[width = 0.48 \textwidth]{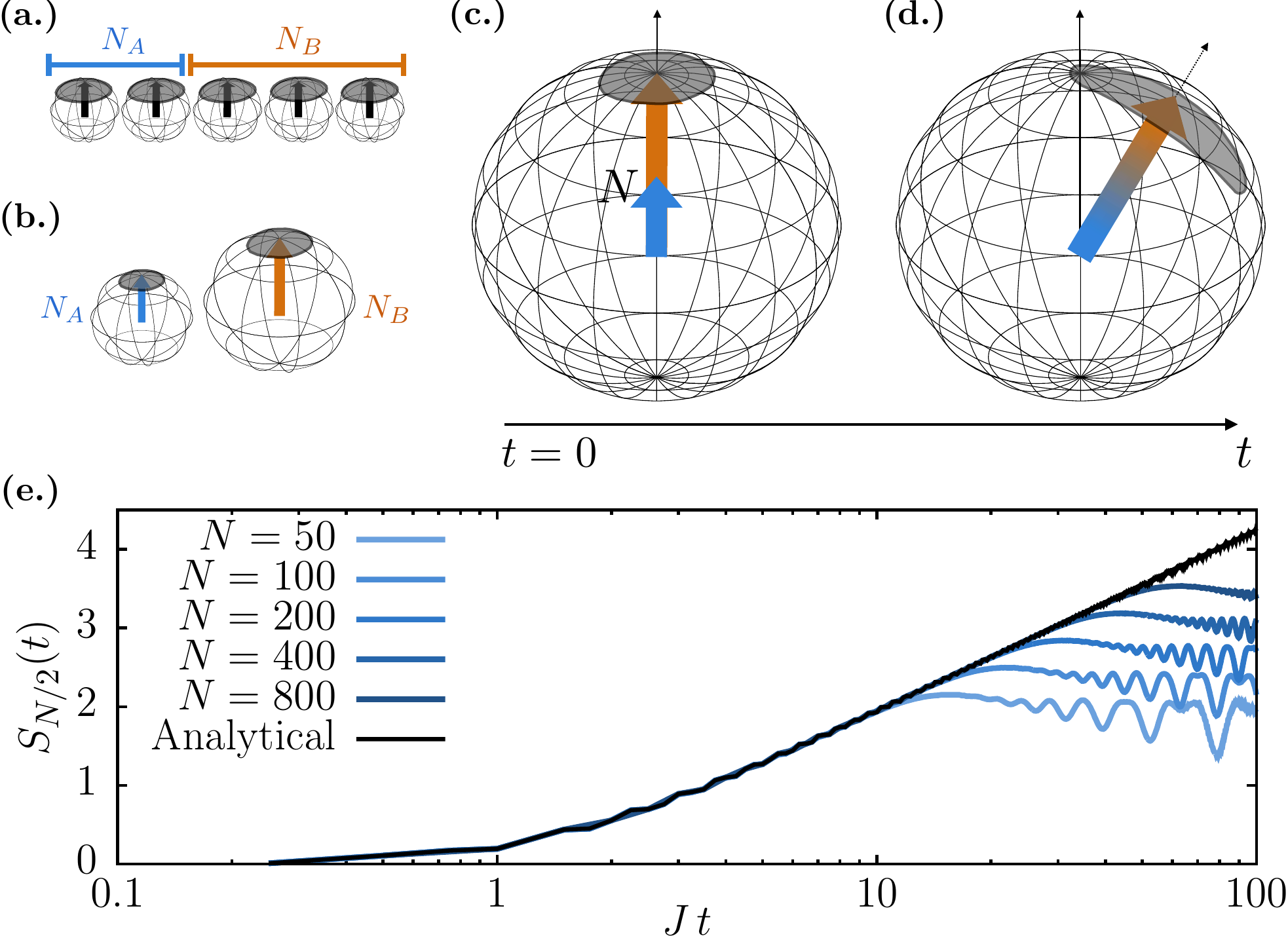}
\vspace{-4mm}
\caption{
Entanglement dynamics and collective spin squeezing in long-range quantum spin systems.
(a) The system is partitioned into two blocks of $N_A$ and $N_B$ spins-$1/2$, initially fully polarized. (b) Collective spins of the two blocks. 
(c) Collective spin in the factorized initial state, represented on a sphere of radius $N/2$. The shaded area represents the quantum uncertainty of  transverse components.
(d) Nonlinear interactions determine spin squeezing, which makes the two blocks increasingly correlated (entangled).
The slow rate of squeezing after non-critical quenches determines the slow growth of entanglement. 
(e) The 
analytical formula for spin-squeezing dynamics derived in this work captures the growth of entanglement entropy until saturation 
 (here, quantum quench 
 in a fully-connected quantum Ising model). 
}
\label{fig:summary}
\end{figure}

\emph{Introduction.}— 
Entanglement entropy 
has been established as a fundamental tool for understanding the out-of-equilibrium dynamics of quantum many-body systems, their thermalization properties,
and the complexity of their numerical simulations \cite{amico2008entanglement, calabrese2004entanglement, CC_QuasiparticlePicture,  alba2018entanglement, 
AlbaThermodynamics,Serbyn2013universal,HuseNandkishore,vidal2003efficient, schollwock2006methods, verstraete2008matrix, banulus2009matrix}.
Significant progress has been achieved for systems with local interactions, where 
the finite speed of correlation spreading generically determines a linear growth in time of bipartite entanglement entropy \cite{calabrese2004entanglement,CC_QuasiparticlePicture,FagottiCalabreseXY,
alba2018entanglement,kimhuse,nahum2017quantum,DeLucaChalker}, unless instances of ergodicity breaking 
{induced by localization or glassy behavior 
cause a distinguished logarithmic slowdown} 
\cite{ProsenMBL, Bardarson2012unbounded, Serbyn2013universal, Cirac:QuasiMBLWithoutDisorder,AbaninProsen:SlowDynamics, SmithEE, SmithLong, brenes2018many}.\\
\indent On the other hand, long-range interacting systems --- ubiquitous in quantum atomic, molecular and optical experiments \cite{RydbergExp1,BernienRydberg,RydbergReviewExp,Browaeys,Albiez2005m, haffner2008quantum, jurcevic2017direct, blatt2012quantum, neyenhuis2017observation, BrittonTrappedIons, RichermeTrappedIons, JurcevicTrappedIons, ZhangDPT, pagano2018cryogenic,PolarMoleculesExp1,PolarMoleculesExp2,MagneticAtoms,DipolarGasExp,black2002Observa,Leroux2010implementa}  
and currently 
the focus of a great attention in quantum many-body theory \cite{GorshkovAreaLawLR,koffel2012entanglement,vodola2014kitaev,frerot2017entanglement,ReyBosonMediated,ReyFossFeigPhononSqueezing,DefenuQuantumONLR,FeySchmidt,Jaschke_2017,VerstraeteQuasiparticlesLR,Defenu2018dynamical,ShammahNoriNumericalDissipativeFullPermutational,
MachadoSlowHeating,SredinickiETHinLR,DelgadoLRsuperconductor,LeroseKapitza,
LukinMBLdipolar,MoessnerRoyMBLinLR,HaukeHeylMBLinLR,BurinMBLinLR,CelardoMBLinLR,MirlinMBLinLR,MorigiMBLinLR,roy2019self,NagGargMBLinLR,ModakMBLinLR,MaksymovMBLinLR,Nandkishore2017many,russomanno2017floquet,PhysRevB.96.104436,PhysRevB.96.134427,RajabpourEEFreeBosonsLR,BrescianiCooperativeShielding,GorshkovConfinement,LeroseDW} 
--- pose a conceptually different and challenging problem. Despite their non-local interactions 
allow {quantum correlations between distant degrees of freedom to build up very quickly} \cite{Gorshkov1,Gorshkov2,MatsutaLRLRbound,ElseLRLRbound,HaukeTagliacozzo,EisertLRCorrelations,
LeporiCorrelationsLR,CarleoLRCorrelations1,CarleoLRCorrelations2,WoutersLRCorrelations,
RoscildeMultispeed}, 
several numerical simulations have shown that entanglement entropy growth after a quench features a dramatic counterintuitive slowdown as the range of interactions is increased: It becomes as slow as logarithmic when the couplings decay algebraically with the distances $r_{ij}$ as $r^{-\alpha}_{ij}$, with 
$\alpha$ smaller than the spatial dimensionality $d$, even in the absence of disorder \cite{Daley,DaleyEssler,pappascrambling}. \\
\indent 
In this work, we {identify the qualitative change in the mechanism which governs the growth of entanglement in  quantum spin systems as the interaction range is increased. 
By means of a systematic 
bosonization of spin excitations, 
we show that the standard quasiparticle contribution to the von Neumann entanglement entropy $S(t)$ is suppressed for $\alpha \le d$ in a prethermal regime}. The growth of $S(t)$ is determined by collective
spin excitations,
directly related to \emph{spin squeezing} \cite{UedaSqueezing,Sorensen1,Sorensen2,TothSqueezing,ReySqueezing,GorshkovSqueezing, pezze2009entanglement}  {for spin-1/2 systems}, as schematically illustrated in Fig.~\ref{fig:summary}.
We demonstrate how a slow \mbox{dynamical rate of spin squeezing after a quench} generically leads to a universal logarithmic growth of $S(t)$ 
as reported in previous numerical studies. The theory further
predicts fast entanglement growth for critical quenches.
Analytical 
 results are supported by numerical simulations 
of long-range quantum Ising chains via exact diagonalization (ED) 
and the matrix-product-state time-dependent variational principle (MPS-TDVP) \cite{Haegeman2016unif, Haegeman2011timedepe}.

\emph{Entanglement entropy in infinite-range spin systems.}— 
{In order to isolate the contribution of collective degrees of freedom to entanglement dynamics,
we start by considering {spin-1/2}  models with full permutational symmetry.}
{In this limiting situation, the only dynamical spin excitations are collective, i.e., uniform, and quasiparticles at all possible non-vanishing momenta are completely suppressed.}
This is realized with arbitrary all-to-all multi-body interactions, described by a Hamiltonian of the form
\beq
\label{eq:Hgeneral}
\hat H = -  \sum_{p=1,2,\dots}  \; \sum_{
\mu_i= x,y,z} \frac{J_{\mu_1 \dots \mu_p}}{N^{p-1}} \sum_{j_1\ne \dots \ne j_p}^N \hat s^{\mu_1}_{j_1} \dots \hat s^{\mu_p}_{j_p}  , 
\eeq
where $\hat{\mathbf{s}}_i$, $i=1,\dots,N$ are quantum spins-$1/2$ (or qubits). The rescaling factor $1/N^{p-1}$ ensures that the energy contribution of all $p$-body interactions is extensive.
These Hamiltonians can be written in terms of the collective spin of the system 
$
\bold{\hat{S}}= \sum_{i=1}^N \bold{\hat{s}}_i
$
only, whose magnitude $\abs{\bold {S}}=\sqrt{S(S+1)}$ is extensive and conserved,
and generically maximal ($S=N/2$) in the ground state \cite{VidalLMGAntiferro},  leading to an effective $\hbar_{\text{eff}} \sim \hbar / N$ \cite{SciollaBiroliMF}. 
The behavior of the system (in and out-of-equilibrium) can thus be understood via systematic semiclassical expansions in quantum fluctuations around the average collective spin
\cite{SM}.\footnote{(This applies to higher spin systems described by Eq. \eqref{eq:Hgeneral}, as well. However, in this case, more general permutationally-symmetric interactions give rise to a description in terms of a larger set of semiclassical variables. See the related comment in the conclusions.)}

We aim at understanding entanglement dynamics in spin systems described by Eq.~\eqref{eq:Hgeneral}. 
For a composite system with Hilbert space $\mathcal H = \mathcal H_A \otimes \mathcal H_B$ in a pure state $\ro = \ket{\psi}\bra{\psi}$, the entanglement between subsystems $A$ and $B$ can be quantified by the Von Neumann entropy 
   ${S_{A} =   - \Tr\big[ \ro_A \log \ro_A \big] } $
   of the reduced density matrix ${\ro_A = \Tr_{B}\,{\hat \rho}}$.
   In this light, we consider a bipartition of the system into two subsystems $A$ and $B$ with $N_A=f_A\, N$ and $N_B=N-N_A=f_B\,N$ spins, respectively. 
The collective spin $\bold {\hat{S}}$ can be correspondingly decomposed as $ \bold {\hat S} = \bold {\hat S}_A + \bold {\hat S}_B $ (see Fig. \ref{fig:summary}).
We extend to the non-equilibrium setting the two-boson formalism developed in Refs.~\onlinecite{vidal2007enta,PhysRevLett.97.220402} (see also Ref. \cite{GroverEEContinuousSB}) for equilibrium problems. 
Preparing the system in a fully polarized state,
both the instantaneous collective spin configuration $\braket{\bold {\hat S}(t)}$ 
 and its quantum fluctuations 
 evolve in time. This can be described by letting the 
$\bold{Z}$-axis follow the evolution of $\braket{\bold {\hat S}(t)}$
\cite{LeroseShort,LeroseLong}.
Quantum fluctuations of the two spins $\bold {\hat S}_{A,B}$ are bosonized via Holstein-Primakoff transformations around their instantaneous average polarization, and  the entanglement between $A$ and $B$ is encoded in the entanglement between these two bosons $(\hat q_{A},\hat p_{A})$ and $(\hat q_{B},\hat p_{B})$.  
The resulting bosonic Hamiltonian governs the dynamics of the transverse quantum fluctuations $(\hat Q, \hat P)$ of the global spin $\bold {\hat S}$ around its direction $\mathbf{Z}(t)$. The latter encodes the bosonic excitations of $\bold {\hat S}_A $, $ \bold {\hat S}_B$ via the relations $\hat Q=\sqrt{f_A } \, \hat q_A + \sqrt{f_B } \, \hat q_B$ and $\hat P=\sqrt{f_A } \, \hat p_A + \sqrt{f_B } \, \hat p_B$.
(See the Supplemental Material~\cite{SM} for the details.) 
We may now safely perform a quadratic approximation, 
as the spin fluctuations are subextensive 
in ground states \footnote{Even at quantum critical points: see, e.g., Refs. \cite{vidalHP, vidal}.} as well as  out of equilibrium until the so-called \emph{Ehrenfest time scale}  --- which diverges with 
$N$, see below. 
{The time-evolution of quantum fluctuations $(\hat Q, \hat P)$ is then formally equal to the classical evolution of displacements away from the trajectory of the initial state under consideration.}

Within this formalism, the Von Neumann entanglement entropy between the two  subsystems $ A$ and $ B$ is computed with standard Gaussian bosonic techniques \cite{SM,BarthelEEquadratic,gaussianQI}, obtaining
 \begin{multline}
\label{eq:EE}
S_A = \sqrt{ 1 + 4 f_A f_B \; \langle  \hat n_{\text{exc}} \rangle } \arccoth \left(\sqrt{ 1 + 4 f_A f_B \; \langle  \hat n_{\text{exc}} \rangle } \right) \\ + \frac{1}{2} \log \big( f_A f_B \; \langle  \hat n_{\text{exc}} \rangle \big),
\end{multline}
where  $\hat n_{\text{exc}} = 
(\hat Q^2 + \hat P^2  -1 )/ 2
$
represents  the number of bosonic excitations of the collective spin $\bold{\hat S}$. 
The function on the right-hand side vanishes for $\langle \hat n_{\text{exc}} \rangle \to 0$ and grows as $\frac{1}{2} \log \langle\hat n_{\text{exc}} \rangle$ for $\langle\hat n_{\text{exc}}\rangle \gg 1$.
Hence, equation \eqref{eq:EE}
clarifies that the state of subsystem $A$ (or $B$) is pure only if $ \langle  \hat n_{\text{exc}} \rangle = 0$, i.e., if the spin system is fully polarized (coherent), as occurs in the absence of interactions. 
{Conversely, the state is entangled in the presence of collective spin excitations.} 
This comes via
\emph{spin squeezing}, a concept first introduced in Ref.~\onlinecite{HiroshimaSqueezing} and usually quantified by the minimal transverse variance of collective spin fluctuations  \cite{WinelandSqueezing,SpinSqueezingReview} as
$
\xi^2  \equiv \Min_{\abs{\mathbf{u}}=1,\mathbf{u}\perp \mathbf{Z}} \Big\langle \big(\mathbf{u}\cdot \bold {\hat S}\big)^2 \Big\rangle / (N/4) 
$. It is
equal to $1$ for coherent states, and smaller for squeezed states (see, e.g., Refs.~\onlinecite{HiroshimaSqueezing,SpinSqueezingReview}). 
At the level of Gaussian fluctuations relevant here, one derives that \cite{SM,gaussianQI}
\beq
\label{eq:squeezing}
\xi^2 = 1+2 \langle  \hat n_{\text{exc}} \rangle - 2 \sqrt{ \langle  \hat n_{\text{exc}}\rangle(1+ \langle  \hat n_{\text{exc}}\rangle)} \; .
\eeq
It has long been known \cite{Sorensen1,Sorensen2, pezze2009entanglement} that collective spin squeezing is a witness of many-body quantum entanglement. {Equations (\ref{eq:EE}) and (\ref{eq:squeezing}) express the quantitative link} --- pictorially illustrated in Fig.~\ref{fig:summary} ---  between the entanglement entropy $S_A$ in collective spin models in and out of equilibrium and the spin-squeezing parameter $\xi$.

\begin{figure}[t]
\fontsize{12}{10}\selectfont
\vspace{-4mm}
\includegraphics[width=0.48\textwidth]{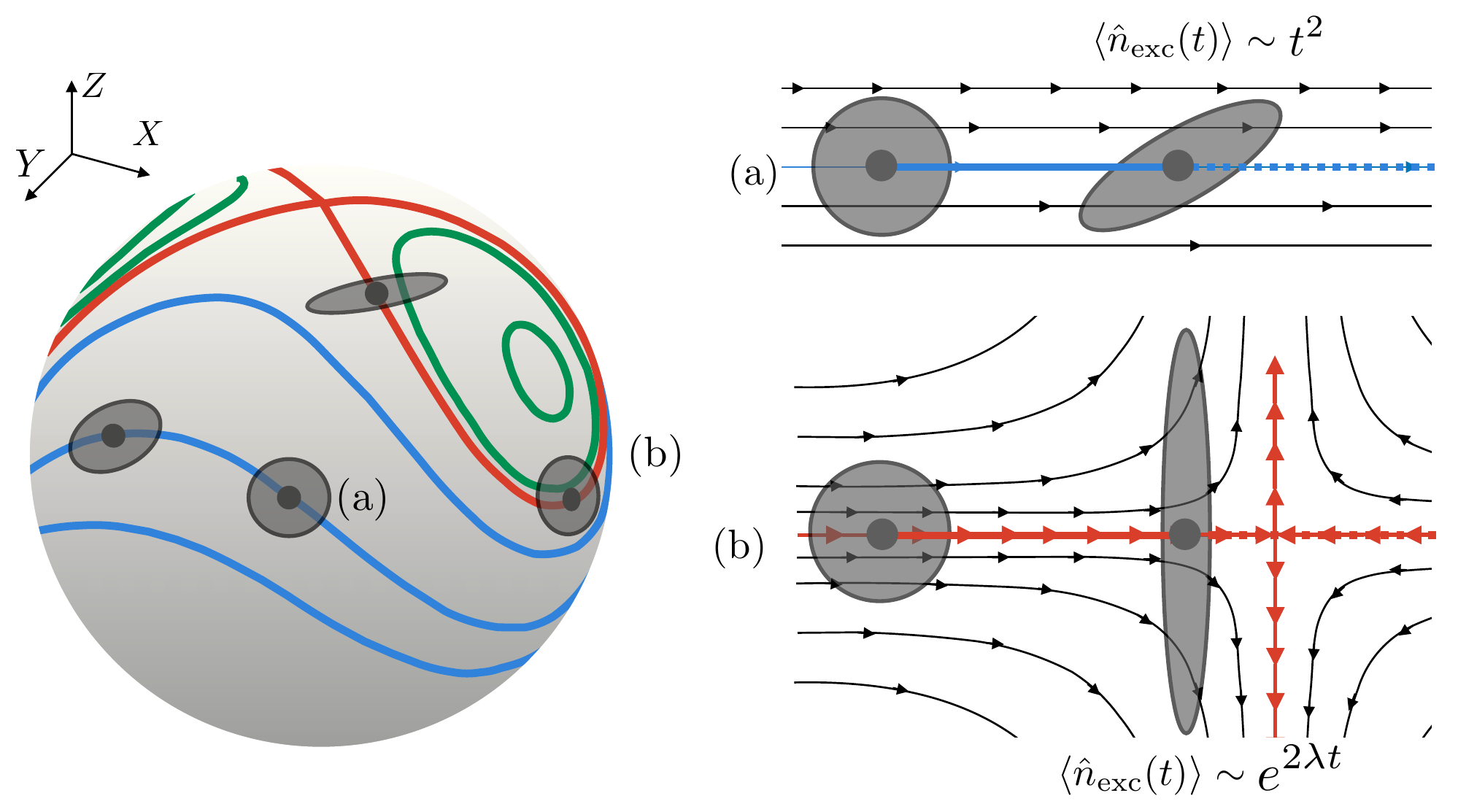}
\vspace{-4mm}
\caption{
Schematic illustration of spin-squeezing dynamics. 
The classical phase space of a ferromagnetic fully-connected Ising model [cf. Eq.~(\ref{eq:Halpha}) with $\alpha=0$]  is pictured on the left. It features ferromagnetic (green) and paramagnetic (blue) periodic trajectories,  separated by a critical trajectory (\emph{separatrix}, red).  
Initial spin-coherent (non-squeezed) states (a) and (b) at $t=0$ are represented by points surrounded by small grey circles representing the quantum fluctuations of transverse spin components. 
Due to nonlinear interactions, the spin state undergoes squeezing, quantified by the parameter $\xi(t)$ [cf. Eq. \eqref{eq:squeezing}]. 
The rate of squeezing is governed by the separation of nearby semiclassical trajectories, and by Eq.~\eqref{eq:EE} it determines the rate of growth of entanglement entropy. 
 Right panels: 
 (a) For generic (non-critical) quenches, nearby trajectories separate linearly in time, leading to a polynomially fast squeezing. 
(b) For a critical quench, the collective spin lies on the stable manifold of an unstable fixed point in phase space. In this case, nearby trajectories separate exponentially fast in time at a rate $\lambda$ set by the eigenvalue of the linearized flow. }
\label{fig:flow}
\end{figure}

{The growth of entanglement entropy out of equilibrium is completely determined by the dynamical generation of the collective excitations $\langle \hat n_{\text{exc}}(t)\rangle$, which is in turn induced by the interactions. 
This rate  depends on the nature of the underlying classical dynamics. 
In fact, the time-dependent quadratic Hamiltonian 
that governs the \emph{quantum} evolution of fluctuations ($\hat Q$,$\hat P$), 
describes as well the \emph{classical} evolution of small displacements away from it in the linear approximation \cite{littlejohn}, see Fig.~\ref{fig:flow}(a) for an illustration. 
%
In classical integrable systems, trajectories originating from nearby initial conditions separate linearly in time, which implies $\hat Q(t), \hat P(t) \sim t$.
Since $ \hat n_{\text{exc}}$ is quadratic in the spin fluctuations $\hat Q$,$\hat P$, the collective excitations grow polynomially in time $\langle \hat n_{\text{exc}}(t)\rangle \sim t^2$ after a generic quench \cite{SciollaBiroliMF}.
This leads to
\beq
S_A(t) \sim \log t
\eeq
by Eq.~\eqref{eq:EE}. 
A more rigorous analysis} \cite{SM} further predicts periodic oscillations of $S_A(t)$ superimposed to the logarithmic growth, the period being that of the underlying classical trajectory. 
In the (non-generic) case of quenches to dynamical critical points \cite{SciollaBiroliMF}, the collective spin lies on unstable trajectories (\emph{separatrices}) in the classical phase space, around which displacements grow exponentially in time, with a rate set by the positive eigenvalue $\lambda$ of the corresponding unstable fixed point on which they terminate, see Fig. \ref{fig:flow}(b). The out-of-equilibrium generation of collective excitations is thus exponentially fast, $\langle \hat n_{\text{exc}}(t) \rangle \sim  e^{2\lambda t}$, leading to a linear growth of entanglement entropy with a predicted slope $S(t) \sim \lambda t$ and without superimposed oscillations.
 (See also the related discussion in Refs. \cite{zurek1995quantum,Bianchi2018,BianchiModakRigolEntanglementBosons}.)
In all cases, entanglement entropy saturates $S_A \sim \log N_A$ at the Ehrenfest time scale defined by ${\langle \hat n_{\text{exc}}(t_{\text{Ehr}}) \rangle \sim N}$. 

\emph{Spatially-decaying interactions.}— 
{We are now in a position to understand the effects  of having slowly-decaying interactions on entanglement dynamics.}
For the sake of definiteness, we focus on periodic $d$-dimensional lattices of spins with arbitrary \emph{two-body} long-range interactions, described by a Hamiltonian as in Eq. \eqref{eq:Hgeneral} with $p\le2$, where now $j=1,\dots,N=L^d$ label lattice sites at positions denoted $\mathbf{r}_{j}$, and the uniform couplings $J_{\mu\nu}$, with $\mu,\nu=x,y,z$, are replaced by  $J_{\mu\nu} / \lvert\mathbf{r}_i-\mathbf{r}_j\rvert^{\alpha}$ 
 \footnote{The distance on the lattice $\lvert\mathbf{r}_i-\mathbf{r}_j\rvert$ is thus taken to be $\sqrt{\sum_{\mu=1}^d \big[ \Min(|r_i^\mu - r_j^\mu|, L-|r_i^\mu - r_j^\mu |)\big]^2}$}.
The exponent $\alpha\ge0$ characterizes the algebraic decay of spin-spin interactions. A Ka{c} rescaling factor  $1/\mathcal{N}_{\alpha,N}$ with $\mathcal{N}_{\alpha,N}=\sum_{ i \ne j } \lvert\mathbf{r}_i-\mathbf{r}_j\rvert^{-\alpha} / N$ replaces the $1/N$ factor in Eq.~\eqref{eq:Hgeneral}, ensuring the extensivity of the Hamiltonian for $\alpha\le d$ \cite{KacNormalization}. The fully-connected limit is recovered by letting $\alpha\to0$.

When interactions decay with the distance between spins, the full permutational symmetry of the infinite-range Hamiltonian \eqref{eq:Hgeneral} is broken and the 
finite-wavelength 
spin modes participate 
in 
the dynamics. These excitations now allow the system to explore the full Hilbert space beyond the Dicke manifold, i.e.,  “inside the Bloch sphere'', and the system may be expected to thermalize by accumulating extensive entanglement entropy \cite{AlbaThermodynamics}. However, {we demonstrate that these quasiparticles are  weakly excited in typical quenches, and hence yield 
only a bounded contribution to entanglement growth for a long  
temporal regime.
In this case, 
the analysis of 
dynamical spin squeezing
captures the leading behavior of $S(t)$ even for $0<\alpha< d$.}

 The two-boson approach to entanglement dynamics described above was strictly based on the conservation of the collective spin magnitude. In order to treat a truly many-body problem, one can refine the approach as follows. 
We rewrite the Hamiltonian in terms of the Fourier spin modes ${\tilde{s}_\mathbf{k}^{x,y,z}= \sum_j e^{-i\mathbf{k}\cdot \mathbf{r}_j} \hat s_j^{x,y,z}}$, with  ${ \mathbf{k}=(2\pi/L)(n_1,\dots,n_d)}$, $n_\mu=0,1,\dots,L-1$ varying in the $d$-dimensional Brillouin zone, 
and bosonize individual spin excitations $\bold{\hat s}_j$ $\to$ ($\hat q_j$,$\hat p_j$) 
around the  instantaneous direction $\mathbf{Z}(t)$ of the collective spin $\langle \mathbf{\hat S}(t) \rangle$ 
via Holstein-Primakoff transformations \cite{LeroseShort,LeroseLong,SM}.
One thus obtains a \textit{spin-wave} expansion 
of the time-dependent frame Hamiltonian ${\widetilde{ H}}(t) = \hat H - \boldsymbol{\omega}(t) \cdot \bold {\hat S}$
in terms of $\tilde{q}_\mathbf{k}= L^{-d/2} \sum_j e^{-i\mathbf{k}\cdot \mathbf{r}_j} \hat q_j$ and $\tilde{p}_\mathbf{k}= L^{-d/2} \sum_j e^{-i\mathbf{k}\cdot \mathbf{r}_j} \hat p_j$ with all possible momenta $\mathbf{k}$, expressed up to $\mathcal{O}(1/\sqrt{N})$ terms as
\footnote{In this case, the dynamical generation of spin waves modify the classical trajectory of the collective spin, due to the feedback from quantum fluctuations. This effect is mostly negligible here, except for critical quenches~\cite{LeroseShort,LeroseLong}.}
\begin{multline}
\label{eq:Hsw}
\widetilde{\hat H}(t) \; \simeq  \;
  \; \widetilde{\hat H}_{0}(t) -  \sum_{\mathbf{k}\neq 0}  \widetilde{f}_{\alpha,\mathbf{k}} \bigg[ 
     J_{qq}(t
     ) \; \frac{\tilde{q}_\mathbf{k} \tilde{q}_{-\mathbf{k}}}{2}  \\ 
     + J_{pp}(t
     )\; \frac{\tilde{p}_\mathbf{k} \tilde{p}_{-\mathbf{k}}}{2} 
      +
        J_{qp}(t
        )\; \frac{\tilde{q}_\mathbf{k} \tilde{p}_{-\mathbf{k}}+\tilde{p}_\mathbf{k} \tilde{q}_{-\mathbf{k}} }{2}
       \bigg] \, , 
\end{multline}
where $\widetilde{f}_{\alpha,\mathbf{k}} = \frac{1}{\mathcal{N}_{\alpha,N}} \sum_{j (\ne i)} \frac{e^{-i\mathbf{k}\cdot (\mathbf{r}_j-\mathbf{r}_i)}}{\lvert\mathbf{r}_j-\mathbf{r}_i\rvert^{\alpha}} $ is proportional to the Fourier transform of the interactions, $J$'s are coefficients depending on the collective spin trajectory,
and the collective-mode Hamiltonian $\widetilde{H}_{0}(t)$
accounts for the infinite-range part 
$\widetilde{f}_{\alpha,\mathbf{0}} \, \delta_{\mathbf{k},\mathbf{0}} \equiv \delta_{\mathbf{k},\mathbf{0}}  $ of the interaction $\widetilde{f}_{\alpha,\mathbf{k}}$.
$\widetilde{H}_{0}(t)$ is independent of $\alpha$ and describes the dynamics of collective spin fluctuations $\hat Q\equiv 
 \tilde{q}_\mathbf{0}$ and $\hat P \equiv 
  \tilde{p}_\mathbf{0}$
as detailed above,  
and conserves the bosonic occupation numbers $
\hat n_{\mathbf{k}\ne\mathbf{0}} \equiv ( \tilde{q}_\mathbf{k} \tilde{q}_{-\mathbf{k}}+\tilde{p}_\mathbf{k} \tilde{p}_{-\mathbf{k}}-1)/2
$ of all the spin-wave modes with finite wavelength \footnote{Note that this is rigorously true to all orders in the Holstein-Primakoff expansion.}.
The dynamical excitation of spin waves with finite wavelengths for $\alpha>0$ is responsible for modifications to the spin-squeezing-induced entanglement entropy growth.
As is evident in Eq.~\eqref{eq:Hsw}, their impact is controlled by the strength of the finite-range part  $ \widetilde{f}_{\alpha,\mathbf{k}\ne\mathbf{0}}$ of the interaction.  
%
In fact, the following estimate can be derived for $\alpha < d$ \cite{SM} 
$
\lvert \widetilde{f}_{\alpha,\mathbf{k}\ne\mathbf{0}}  \rvert \, \le \, 
\text{const} \times  \frac{1}{(\abs{\mathbf{k}} L)^{\beta}} $, with $\beta\equiv\Min\big(d-\alpha,(d+1)/2\big)
$
(for $\alpha=d$ the power law is replaced by a logarithm).
This bound implies that for all fixed $\mathbf{k}\ne\mathbf{0}$, 
the coupling $\widetilde{f}_{\alpha,\mathbf{k}}$ is vanishingly small in thermodynamic limit $L\to\infty$ whenever $\alpha\le d$, and hence the associated number of bosonic excitations is an approximate constant of motion,
\beq
\label{eq:bound}
\Big\lvert \Big\langle \big[\hat n_{\mathbf{k}\ne\mathbf{0}}, \widetilde{\hat H}(t) \big]  \Big\rangle \Big\rvert
\le \; \frac{ \text{const} }{(\abs{\mathbf{k}} L)^{\beta}}.
\eeq
Therefore, there exists a long time scale $T_{\text{sw}}\sim N^{\beta/d}$, during which the dynamical excitation of spin waves with finite wavelengths is suppressed \footnote{Note that this justifies a posteriori the Holstein-Primakoff approach, as the density of spin waves  remains small over a long time window} (note the interesting relation to the \emph{prethermalization} time in Ref.~\cite{mori2018prethermalization}). 
{On the other hand, permutational symmetry may  severely break over large length scales 
via excitations with $\abs{\bold k}  \propto 1/L$. 
Their dynamics governed by the Hamiltonian \eqref{eq:Hsw}, is equivalent to a \emph{discrete} set of periodically driven quantum 
oscillators, the drive being induced by the precession of the collective spin.
From a stability analysis, we find that for typical quenches these long-wavelength spin-wave modes are non-resonantly driven and hence weakly excited. 
Their resulting contribution to entanglement dynamics amounts to bounded oscillations on top of the dominant, spin-squeezing induced logarithmic growth. 
Near dynamical criticality, however, resonant excitation of these modes may lead to exponentially growing quantum fluctuations (cf. Ref. \cite{mori2018prethermalization}) and hence linear increase of the entanglement entropy  (see the Supplemental Material \cite{SM} for details).}
We thus conclude that long-range interacting {spin-1/2} systems with $\alpha < d$ typically exhibit logarithmic growth of entanglement entropy. 

\begin{figure}[t]
\fontsize{13.5}{10}\selectfont
\includegraphics[scale = 0.66]{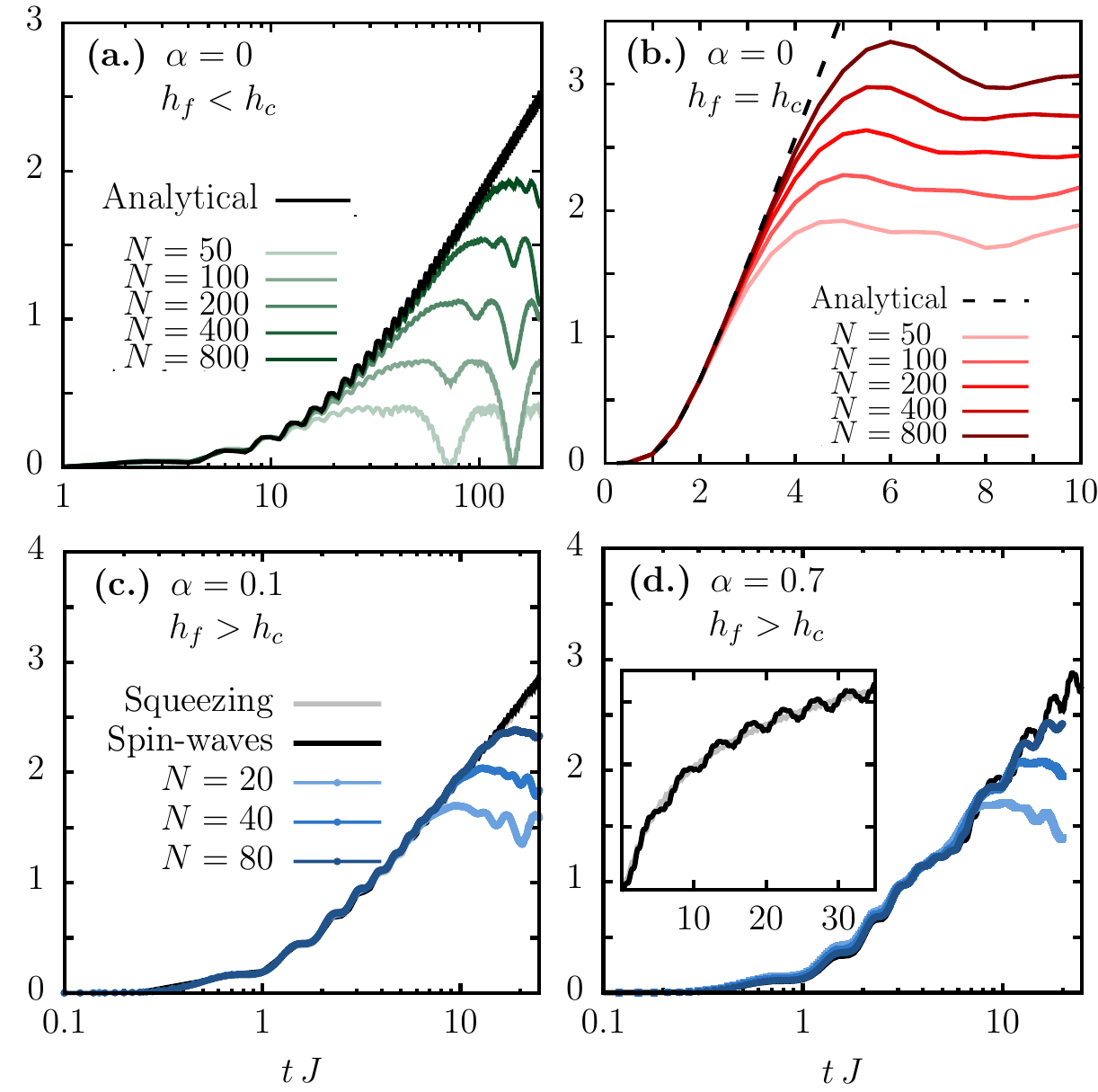}
\vspace{-0mm}
\caption{
Comparison between theory and numerical results for long-range quantum Ising chains. 
Quenches in the transverse field $h_0=0 \to h_f>0$ are considered, and the evolution of the half-system entanglement entropy $S_{N/2}(t)$ is shown.
Top: Fully-connected limit with $\alpha=0$. Analytical results (black lines) are compared with ED data for increasing system sizes $N=20\div 800$. (a.) For a non-critical quench $h_0=0 \to h_f=0.2J$,  the growth of $S_{N/2}(t)$ is  logarithmic up to saturation around $1/2 \log N$ at $t_{ \text{Ehr}}\sim \sqrt N$ (b.) For the critical quench $h_f=h_c=J/2$, the growth of $S_{N/2}(t)$ is linear until ${t_{ \text{Ehr}}\sim \log N}$, with a slope $\lambda_{h_c}=J$.
Bottom: Deep quench with $h_f=2J$ in long-range interacting chains with $\alpha > 0$. The contribution due to collective spin squeezing [Eq. \eqref{eq:EE}] {and the full spin-wave calculation of the time-dependent entanglement (see the main text)} are compared with MPS-TDVP data for $N=20\div 80$ converged with bond dimension $D=128$, for  $\alpha=0.1$
(c.) and $\alpha=0.7$ (d). As $\alpha$  increases, finite corrections due to long-wavelength spin waves appear on top of the dominant spin-squeezing-induced logarithmic growth, {see the inset}. 
}
\label{fig:3}
\end{figure}
\emph{Numerical simulations.}— 
We test  our analytical understanding 
in paradigmatic one-dimensional long-range quantum Ising chains, described by the Hamiltonian
\beq
\label{eq:Halpha}
\hat H = 
 -  \frac{J}{\mathcal{N}_{\alpha,N}}  \sum_{ i < j }^N \frac{\hat \sigma_{i}^x \hat \sigma_{j}^x}{\lvert i-j\rvert^{\alpha}}
 - h \sum_{i}^N \hat \sigma_{i}^z ,
\eeq
where $i,j=1,\dots,N$, $\hat \sigma_i^{x,z}$ are Pauli matrices, $h$ is a global transverse magnetic field and $\mathcal{N}_{\alpha,N}$ is the Ka{c} rescaling factor introduced above. 
%

We compare the numerical computations of entanglement entropy evolution at finite $N$ with the analytical calculation of the spin-squeezing  contribution  [Eq.~(\ref{eq:EE})] 
{and with the full spin-wave calculation, 
obtained from Eq. \eqref{eq:Hsw} via standard bosonic techniques \cite{BarthelEEquadratic,gaussianQI,SM,BianchiModakRigolEntanglementBosons}.}
For the sake of illustration, we focus here on the initial state $\ket{\psi_0}=\ket{\rightarrow \, \rightarrow \, \dots \rightarrow \,}$, i.e., on quenches in the transverse field from $h_0=0$ to $h_f$.
As Figs.~\ref{fig:summary} and \ref{fig:3} show, in all cases the numerical data are captured by the corresponding analytical curves 
for 
$t \lesssim t_{ \text{Ehr}}(N)$.
In the fully-connected limit $\alpha\to0$, equivalent to the Lipkin-Meshov-Glick model \cite{LMGref}, Eq. (\ref{eq:EE}) is exact in the thermodynamic limit and the finite-size 
ED data 
perfectly match it 
before saturation at the Ehrenfest time, $t_{ \text{Ehr}} \sim \sqrt{N}$ for generic quenches [$h_0=0 \to h_f = 2J$ in Fig.~\ref{fig:summary} (bottom) and $h_f=0.2 J$ in  Fig.~\ref{fig:3}(a)] and $t_{ \text{Ehr}} \sim \log{N}$ for the critical quench [$h_f = h_c \equiv J/2$ in  Fig. \ref{fig:3}(b), cf. the red line in  Fig.~\ref{fig:flow}(b)], corresponding to the \emph{dynamical phase transition} of the model  \cite{SciollaBiroliMF, ZunkovicMF, Zunkovic2016, ZhangDPT}. 
For spatially-decaying interactions with $0<\alpha<1$, we employ the MPS-TDVP 
 \cite{Haegeman2016unif, Haegeman2011timedepe} {with periodic boundary conditions 
 (see the Supplemental Material \cite{SM} for details). 
Upon increasing $N$, the TDVP data approach the full spin-wave entanglement entropy, for all considered values of $\alpha$ and quench parameters, as shown in the examples in Fig.~\ref{fig:3}(c),(d).
This analysis confirms that the growth of $S(t)$ is logarithmic for typical initial configurations. 
} 
%
For further discussion, including varying initial states, bipartition sizes and {details on the spin-wave analysis,} see the Supplemental Material \cite{SM}. 

\emph{Outlook.}— 
The mechanism 
unveiled in this work complements the available paradigms for entanglement dynamics characterizing systems with local interactions \cite{calabrese2004entanglement,CC_QuasiparticlePicture,FagottiCalabreseXY,
alba2018entanglement,nahum2017quantum, kimhuse,DeLucaChalker}
and, at the same time,
improves the current understanding of the
efficiency of ``classical'' simulations of quantum long-range interacting {spin} chains with matrix-product-state techniques \cite{Haegeman2016unif, Haegeman2011timedepe}.
Similar approaches can be used to characterize entanglement entropy dynamics in more general many-body systems \cite{SciollaBiroliMF}, {including higher-spin systems, whose semiclassical dynamics may display chaos and hence fast entanglement growth for generic quenches.}
In connection with known instances of ergodicity breaking stemming from long-range interactions \cite{RuffoLRErgodicityBreaking, GongPrethermalizationLR,neyenhuis2017observation, mori2018prethermalization,GorshkovConfinement, LeroseDW, koffel2012entanglement, vodola2014kitaev, Daley,DaleyEssler, Zunkovic2016, HalimehDPT}, 
the suppression of spin-wave excitations 
clarifies the role of long-range interactions in constraining quantum dynamics 
to small portions of the many-body Hilbert space {in the relevant time regime.}
Concerning the slow entanglement growth, this is reminiscent of localized or glassy dynamics  \cite{Cirac:QuasiMBLWithoutDisorder,AbaninProsen:SlowDynamics,SmithEE,SmithLong, brenes2018many}. 

The bridge 
between bipartite entanglement entropy and other measures of multi-particle entanglement, most notably spin squeezing \cite{UedaSqueezing,Sorensen1, Sorensen2, TothSqueezing,ReySqueezing,GorshkovSqueezing,pezze2009entanglement,hyllus2012qfi,strobel2014fisher}, may help to elucidate the relation between bipartite and multipartite entanglement, making the latter 
an equivalent metrologically useful  measure for systems with long-range interactions. Most importantly, this connection 
has direct experimental relevance 
for the detection of entanglement and its dynamics via measurements of collective quantities \cite{RefaelSilvaEntanglement,JessenKickedTop}, 
the experimental accessibility of which is well-established with standard techniques and tools of quantum atomic experiments, ranging from spinor condensates and cavity-QED systems to 
trapped ions \cite{strobel2014fisher, bohnet2016quantum, garttner2017measuring}. \\

\begin{acknowledgments} 
We gratefully thank P. Calabrese, M. Dalmonte, R. Fazio and  J.~Goold for insightful comments on the manuscript, G. Giudici for sharing with us his TDVP code, and A. De Luca, J. De Nardis, A. V. Gorshkov, P. Hauke, J. G. Hirsch, J. Marino, G. Morigi, C. Murthy,  G. Pagano, S. Parameswaran, A. M. Rey, L. F. Santos, A. Silva, P. Titum, Q. Zhuang and B. \v{Z}unkovi\v{c} for interesting discussions  on the subject matter of this work.
\end{acknowledgments}

\bibliography{biblio2}  

\end{document}


\title{
Supplemental Material \\ Origin of the slow growth of entanglement in long-range interacting spin systems} 

\author{Alessio Lerose} 
\affiliation{SISSA --- International School for Advanced Studies, via Bonomea 265, I-34136 Trieste, Italy}
\affiliation{INFN --- Istituto Nazionale di Fisica Nucleare, Sezione di Trieste, I-34136 Trieste, Italy}

\author{Silvia Pappalardi} 
\affiliation{SISSA --- International School for Advanced Studies, via Bonomea 265, I-34136 Trieste, Italy}
\affiliation{Abdus Salam ICTP --- International Center for Theoretical Physics, Strada Costiera 11, I-34151 Trieste, Italy}

\date{\today} 



\maketitle
In this Supplemental Material, we provide additional information on the analytical calculations and numerical results presented in the Letter. In Sec. \ref{sec:holsta_prima} we illustrate the time-dependent Holstein-Primakoff transformations and show how, in infinite-range models, the dynamics of subsystem fluctuations is determined by collective fluctuations. In Sec. \ref{app:enta_colle_exc}, the results presented in the Letter are derived, concerning von Neumann entanglement entropy and its link to spin squeezing and to the dimensionless effective temperature. In Sec. \ref{sec:neqgrowth} we rigorously demonstrate  the polynomial growth of $\langle \hat n_{\text{exc}}(t)\rangle$ and discuss the case of critical quenches. Section \ref{sec:saturation} is devoted to a brief comment on the entanglement entropy saturation. In Sec. \ref{eq:alphaneq0}, we provide the details about the discussion on spatially-decaying interactions. We conclude in Sec. \ref{sec:LMG} by analyzing the long-range quantum Ising chain  and by giving  the  details of our numerical computations.

\section{Two-boson formalism in and out-of-equilibrium}

\label{sec:holsta_prima}

We consider a partitioning --- pictorially illustrated in Fig.~1 of the Letter --- of a system described by Eq. (1) of the Letter, 
 into two subsets of spins $A$ and $B$ with $N_A$ and $N_B=N-N_A$ spins, respectively. The collective spin $\bold {\hat{S}}$ can be decomposed as 
\beq
\bold {\hat S} = \bold {\hat S}_A + \bold {\hat S}_B.
\eeq
Concerning typical states  in and out of equilibrium characterized by a maximal collective spin magnitude (see the Letter), following Refs.~\onlinecite{vidal2007enta,PhysRevLett.97.220402},  the quantum correlations between subsystems $A$ and $B$ 
can be understood by expanding the two spins $\bold {\hat S}_A$, $\bold {\hat S}_B$ in quantum fluctuations around the direction $\mathbf{Z}=(\sin\theta\cos\phi,\sin\theta\sin\phi,\cos\theta)$ of  $\braket{\bold {\hat S}}$. This is obtained by means of Holstein-Primakoff transformations from spin to canonical bosonic operators (see, e.g., Ref.~\onlinecite{wannier}), expressed by
\begin{equation}
\label{eq:approxH-P}
\left\{
\begin{split}
\hat S^X_{A,B} &= \sqrt{N_{A,B} \, s} \;\; \hat q_{A,B}  \;  + \mathcal{O}\Big(1/\sqrt{N_{A,B}}\Big) \, , \\
\hat S^Y_{A,B}  &= \sqrt{N_{A,B} \, s} \;\; \hat p_{A,B} \;  + \mathcal{O}\Big(1/\sqrt{N_{A,B}}\Big) \, , \\
\hat S^Z_{A,B}  &= N_{A,B} \, s - \hat n_{A,B} \equiv N_{A,B} \,  s - \frac{ \hat q_{A,B}^2+\hat p_{A,B}^2-1}{2},
\end{split} 
\right.
\end{equation}
where $s=1/2$ here, and the rotated frame 
\begin{equation}
\label{eq:newbasis}
\bold{X} \equiv 
\left( \begin{matrix}
\cos\theta \cos\phi \\
\cos\theta\sin\phi \\
-\sin\theta
\end{matrix} \right) ,  \,\,
\bold{Y} \equiv 
\left( \begin{matrix}
-\sin\phi \\
\cos\phi \\
0 
\end{matrix} \right), \,\,
\bold{Z} \equiv 
\left( \begin{matrix}
\sin\theta \cos\phi \\
\sin\theta\sin\phi \\
\cos\theta
\end{matrix} \right) 
\end{equation}
will be determined in such a way that the $Z$-axis is aligned with $\braket{\bold {\hat S}}$, both in and out of equilibrium.
The two bosonic modes $(\hat q_A,\hat p_A)$ and $(\hat q_B,\hat p_B)$ describe spin excitations localized in subsystem $A$ and $B$, respectively. In terms of these two modes, the collective spin of the global system reads
\begin{equation}
\label{eq:approxH-Pcollective}
\left\{
\begin{split}
\hat S^X &= \sqrt{N \, s} \Big( \sqrt{f_A} \; \hat q_{A} + \sqrt{f_B} \; \hat q_{B} \Big)  \; + \mathcal{O}\Big(1/\sqrt{N}\Big) \, , \\
\hat S^Y  &= \sqrt{N \, s} \Big( \sqrt{f_A} \; \hat p_{A} + \sqrt{f_B} \; \hat p_{B} \Big) \;  +  \mathcal{O}\Big(1/\sqrt{N}\Big) \, , \\
\hat S^Z  &= N \, s - \hat n_A - \hat n_B ,
\end{split} 
\right.
\end{equation}
where $f_{A,B} \equiv N_{A,B}/N$ represent the fraction of spins in subsystems $A$ and $B$, respectively (so $f_A+f_B=1$).

We can now express the Hamiltonian in Eq.~(1) of the Letter in terms of the rotated collective spin components $\hat S^X$, $\hat S^Y$, $\hat S^Z$ through the change of frame \eqref{eq:newbasis}, 
and hence systematically expand them in bosonic excitations 
$(\hat q_A,\hat p_A)$, $(\hat q_B,\hat p_B)$ via Eqs.~\eqref{eq:approxH-Pcollective}. 
In this two-boson description, 
the entanglement between subsystems $A$ and $B$ is encoded by the entanglement between these two bosonic modes. 
%
In order to study static and dynamical properties, it is convenient to work with the ``collective'' and ``spin-wave'' modes $(\hat Q,\hat P)$ and $(\hat q,\hat p)$ defined by the following canonical transformation,
\begin{eqnarray}
\label{eq:Qq}
\left\{
\begin{split}
\hat Q &= +\sqrt{f_A} \; \hat q_A + \sqrt{f_B} \; \hat q_B \\
\hat q &= -\sqrt{f_B} \; \hat q_A + \sqrt{f_A} \; \hat q_B \\
\end{split} 
\right. \; ,
\\
\label{eq:Pp}
\left\{
\begin{split}
\hat P &= +\sqrt{f_A} \; \hat p_A + \sqrt{f_B} \; \hat p_B \\
\hat p &= -\sqrt{f_B} \; \hat p_A + \sqrt{f_A} \; \hat p_B \\
\end{split} 
\right. \; ,
\end{eqnarray}
i.e., by a rotation by an angle $\zeta=\arctan(\sqrt{f_B/f_A})$ in the $\hat q_A$-$\hat q_B$ and $\hat p_A$-$\hat p_B$ planes. 
By Eqs.~\eqref{eq:approxH-Pcollective}, in terms of these bosonic modes 
the collective spin reads
\begin{equation}
\label{eq:approxH-PQP}
\left\{
\begin{split}
\hat S^X &= \sqrt{N \, s} \; \hat Q  \; +  \mathcal{O}\Big(1/\sqrt{N}\Big) \, , \\
\hat S^Y  &= \sqrt{N \, s} \; \hat P  \; +  \mathcal{O}\Big(1/\sqrt{N}\Big) \, , \\
\hat S^Z  &= N \, s - \frac{\hat Q^2 + \hat P^2 -1}{2} - \frac{\hat q^2 + \hat p^2 -1}{2} 
\equiv N \, s - \hat n_{\text{exc}} - \hat n_{\text{sw}},
\end{split} 
\right.
\end{equation}
The $(\hat Q,\hat P)$ mode represents uniform collective spin excitations in the entire system, and, accordingly, does not affect the collective spin magnitude $\lvert \hat{\mathbf{S}} \rvert^2$. 
On the other hand, the $(\hat q,\hat p)$ mode represents out-of-phase excitations of spins in subsystems $A$ and $B$, or ``spin waves'', which decrease the collective spin magnitude $\lvert \hat{\mathbf{S}} \rvert^2 =(N/2 -  \hat n_{\text{sw}} )(N/2 -  \hat n_{\text{sw}} +1)$. 

Using Eqs.~\eqref{eq:approxH-P}, the expansion of the Hamiltonian (1) of the Letter 
has the form
\beq
\label{eq:HQP}
\begin{split}
\hat H =&  \; N \; \mathcal{E}_{\text{cl}}(\theta,\phi) +
 \sqrt{N} \Big[ h^{(1)}_Q(\theta,\phi) \; \hat Q
+  h^{(1)}_P(\theta,\phi) \; \hat P \Big] 
\\& 
+ h^{(2)}_{QQ}(\theta,\phi) \; \frac{\hat Q^2}{2} + h^{(2)}_{PP}(\theta,\phi) \; \frac{\hat P^2}{2}
+ h^{(2)}_{QP}(\theta,\phi) \; \frac{\hat Q\hat P+\hat P\hat Q}{2} 
+ h^{(2)}_{\text{sw}}(\theta,\phi) \; \frac{\hat q^2+\hat p^2-1}{2} 
+ \mathcal{O}\Big(1/\sqrt{N}\Big),
\end{split}
\eeq
where the explicit expression of the coefficients $\mathcal{E}_{\text{cl}}$, $h^{(1)}$, $h^{(2)}$ is determined by the couplings $J$ in the Hamiltonian (1) of the Letter. 
The angles $\theta$ and $\phi$ are fixed in such a way that the linear terms $h^{(1)}$ vanish, and 
this yields the minimum point of the classical Hamiltonian $\mathcal{H}_{\text{cl}}$. 
The mode $(\hat q,\hat p)$ enters Eq.~\eqref{eq:HQP} only through the number of bosons $\hat n_{\text{sw}}= (\hat q^2+\hat p^2-1)/2$, and, accordingly, these spin waves cannot be excited in the ground state nor dynamically in infinite-range systems, i.e.,
\beq
\braket{\hat n_{\text{sw}}} \equiv 0 
\eeq
For the same reason one also has vanishing mixed correlations
\beq
\braket{\hat q \hat Q} = \braket{\hat p \hat P} = \braket{\hat q \hat P} = \braket{\hat p \hat Q} \equiv 0.
\eeq
On the other hand, the number of collective excitations $\hat n_{\text{exc}}= (\hat Q^2+\hat P^2-1)/2$ is not conserved. In the ground state, correlations of $Q$ and $P$ 
can be obtained by diagonalizing $\hat H$ with a generalized Bogolubov transformation, 
\beq
\left\{
\begin{split}
\hat Q =& \; e^{\gamma}\cos\eta \; {\hat Q}^\prime - e^{\gamma}\sin\eta \; {\hat P}^\prime \\
\hat P =& \; e^{-\gamma}\sin\eta \; {\hat Q}^\prime + e^{-\gamma}\cos\eta \; {\hat P}^\prime 
\end{split}
\right.
\eeq
with 
\beq
\tan (2\eta ) \equiv  \frac{h^{(2)}_{QP}}{h^{(2)}_{QQ}+h^{(2)}_{PP}} \ ,
\quad \quad \quad
e^{2\gamma} \equiv \frac{
h^{(2)}_{QQ}+h^{(2)}_{PP} - \Delta
}{
h^{(2)}_{QQ}+h^{(2)}_{PP} + \Delta
},
\eeq
and with $\Delta \equiv  (h^{(2)}_{QQ}-h^{(2)}_{PP}) \cos(2\eta) + 2 h^{(2)}_{QP} \sin(2\eta)$.
One finds in fact
\beq
\label{eq:GScorrelations}
\left\{
\begin{split}
G^{QQ} & \equiv\langle\hat  Q^2 \rangle_{\text{GS}} = \frac{1}{2} \big[ \cosh(2\gamma) + \sinh(2\gamma) \cos(2\eta) \big], \\
G^{PP} &\equiv\langle \hat P^2 \rangle_{\text{GS}} = \frac{1}{2} \big[ \cosh(2\gamma) - \sinh(2\gamma) \cos(2\eta) \big], \\
G^{QP}& \equiv\frac{\langle \hat Q\hat P+\hat P\hat Q \rangle_{\text{GS}}}{2} = \frac{1}{2} \sinh(2\gamma) \sin(2\eta) .\\
\end{split}
\right.
\eeq
In particular, 
\beq
\label{eq:GSexcitations}
\langle \hat n_{\text{exc}} \rangle_{\text{GS}} = \frac{G^{QQ}+G^{PP}-1}{2} = \frac{\cosh(2\gamma)-1}{2}.
\eeq



When the system is driven out of equilibrium by varying in time some parameter in the Hamiltonian, one has that both the direction of the collective spin configuration $\theta(t)$, $\phi(t)$ 
 and the collective spin excitations around it $G^{QQ}(t)$, $G^{QP}(t)$, $G^{PP}(t)$ evolve in time.
 Following Refs.~\onlinecite{LeroseShort,LeroseLong},
the motion of the angles $\theta(t)$, $\phi(t)$ can be accounted for by letting the rotated frame $(\mathbf{X},\mathbf{Y},\mathbf{Z})$ in Eq.~\eqref{eq:newbasis} change in time in such a way that the $\bold Z$-axis self-consistently follows the evolution of $\braket{\bold {\hat S}(t)}$. The modified Hamiltonian in this time-dependent frame includes the inertial forces due to the motion of the frame,
and reads
\beq
\label{eq:Hrotatingframe}
\widetilde{\hat H}(t)= \hat H - \boldsymbol{\omega}(t) \cdot \bold {\hat S}
\eeq
with $\omega^X = -\sin\theta \; \dot{\phi}$, $\omega^Y  =  \dot{\theta}$,  and $\omega^Z = \cos\theta \; \dot{\phi}$.
The evolution of $\theta(t)$ and $\phi(t)$ is determined by the vanishing of the linear terms in $\hat S^X$ and $\hat S^Y$, and 
this yields the classical trajectory 
governed by $\mathcal{H}_{\text{cl}}$. 

The resulting time-dependent quadratic part of $\widetilde{\hat H}(t)$, denoted $\widetilde{h}^{(2)}(t)$ and given by
\beq
\label{eq:H2tilde}
\begin{split}
\widetilde{h}^{(2)}_{QQ, PP,\text{sw}}(t) & \equiv h^{(2)}_{QQ,PP,\text{sw}}\big(\theta(t),\phi(t)\big) - \cos\theta(t) \; \dot{\phi}(t), \\
\widetilde{h}^{(2)}_{QP}(t) & \equiv h^{(2)}_{QP}\big(\theta(t),\phi(t)\big) ,
\end{split}
\eeq
[cf. Eq.~\eqref{eq:HQP}]
 determines the dynamical generation of collective bosonic excitations $(\hat Q,\hat P)$, which can be monitored thorugh the correlation functions $G^{QQ}(t)$, $G^{QP}(t)$, $G^{PP}(t)$. In order to compute them, one starts from the Heisenberg equations of motion
\beq
\label{eq:eom}
\left\{
\begin{split}
\dot{\hat Q}=& +\widetilde{h}^{(2)}_{QP}(t)  \; \hat Q + \widetilde{h}^{(2)}_{PP}(t)  \; \hat P,\\
\dot{\hat P}= & - \widetilde{h}^{(2)}_{QQ}(t)  \; \hat Q - \widetilde{h}^{(2)}_{QP}(t)  \; \hat P.
\end{split}
\right.
\eeq
Denoting the solution
\beq
\label{eq:U(t)}
\begin{pmatrix}
\hat Q(t) \\
\hat P(t)
\end{pmatrix}
= U(t) 
\begin{pmatrix}
\hat Q(0) \\
\hat P(0)
\end{pmatrix}
\eeq
and collecting the dynamical correlations
\beq
\label{eq:collective_dyn_corr}
\left\{
\begin{split}
G^{QQ}(t) & \equiv\langle \hat Q^2(t) \rangle , \\
G^{PP}(t) &\equiv\langle \hat P^2(t) \rangle , \\
G^{QP}(t)& \equiv\frac{\langle \hat Q(t)\hat P(t)+\hat P(t)\hat Q(t) \rangle}{2} ,\\
\end{split}
\right.
\eeq
in the matrix 
\beq
\label{eq:G(t)}
G(t)
=  
\begin{pmatrix}
G^{QQ}(t) & G^{QP}(t) \\
G^{QP}(t) & G^{PP}(t) 
\end{pmatrix}
= U(t) \, G(t=0) \, U^T(t),
\eeq
the number of dynamically generated excitations can be expressed as
\beq
\label{eq:ne}
\langle \hat n_{\text{exc}}(t) \rangle = \frac{G^{QQ}(t) + G^{PP}(t) - 1}{2} = \frac{1}{2} \Tr \bigg[  G(t) - \frac{\mathbb{1}}{2} \bigg].
\eeq
Note that $\det G(t) \equiv 1/4$, which is an exact property of \emph{pure} Gaussian states [cf. Eq.~\eqref{eq:GScorrelations}] preserved by the evolution Eq.~\eqref{eq:eom}.

The two-boson formalism outlined in this Section relies on the truncation of the Holstein-Primakoff transformation (\ref{eq:approxH-P}), which is accurate for Gaussian states with a small number of collective excitations $\langle \hat n_{\text{exc}} \rangle \ll N$ compared to the system size. This assumption is generically valid for ground states, even at quantum critical points \cite{vidal,vidalHP}, as well as in the non-equilibrium setting up to a time scale which diverges with the system size (see below).


\section{Entanglement entropy, collective excitations and spin squeezing}
\label{app:enta_colle_exc}



For a given stationary or time-evolving state of the system, 
we can compute the entanglement entropy between the two  subsystems $ A$ and $ B$. This amounts to computing the entanglement entropy between the two bosons $(\hat q_A,\hat p_A)$ and $(\hat q_B,\hat p_B)$, corresponding to spin excitations localized in $A$ and $B$ respectively.
The reduced density matrix of subsystem $A$ is a Gaussian state of the boson $(\hat q_A,\hat p_A)$ completely determined by the correlation matrix
\beq
\begin{split}
G_A&=
\begin{pmatrix}
\langle \hat q_A^2 \rangle & \frac{\langle \hat q_A\hat p_A+\hat p_A \hat q_A \rangle}{2} \\
\frac{\langle \hat q_A\hat p_A+\hat p_A \hat q_A \rangle}{2} & \langle \hat p_A^2 \rangle 
\end{pmatrix}
\equiv
\begin{pmatrix}
G^{q_Aq_A} & G^{q_Ap_A} \\
G^{q_Ap_A} & G^{p_Ap_A} 
\end{pmatrix}.
\end{split}
\eeq 
The Von Neumann entropy 
 of a single boson $(\hat q_A, \hat p_A)$ in such a Gaussian state can be expressed in terms of the determinant of $G_A$ as \cite{BarthelEEquadratic}
\begin{equation}
\label{eq:EE}
S_A = 2 \sqrt{ \det G_A } \arccoth \left(2 \sqrt{ \det G_A } \right) + \frac{1}{2} \log \left( \det G_A - \frac{1}{4}\right).
\end{equation}
On the other hand, the matrix $G_A$ can be easily related to the correlation matrix  $G$ of collective excitations $(Q,P)$ in the system by inverting Eqs.~\eqref{eq:Qq}-\eqref{eq:Pp}. The explicit computation shows that its determinant amounts to
\beq
\label{eq:EEandnexc}
\det G_A = \frac{1}{4} + \frac{1}{2} f_A f_B \Tr\bigg(G-\frac{\mathbb{1}}{2}\bigg) 
= \frac{1}{4} + f_A f_B \; \langle  \hat n_{\text{exc}} \rangle
\eeq
Equations (\ref{eq:EE}) and (\ref{eq:EEandnexc}) complemented by the evolution in time of $\langle \hat n_{\text{exc}}\rangle$ connect entanglement entropy dynamics to the growth of the quantum fluctuations of the collective spin, whose determination represents the main result of this Section. 
Taking the limits of small or large $\langle  \hat n_{\text{exc}}\rangle$ in Eqs.~\eqref{eq:EE}, \eqref{eq:EEandnexc}, one finds 
 \begin{subequations}
\begin{align}
 S_A & \sim  f_A f_B \, \langle  \hat n_{\text{exc}} \rangle \,  \log  f_A f_B \langle  \hat n_{\text{exc}} \rangle 
 \quad \quad \quad\;\;\, \text{for} \quad \quad \langle  \hat n_{\text{exc}} \rangle\ll 1 \ , \\
 \label{eq:big_S}
 S_A & \sim  1+\frac{1}{2} \log  f_A f_B + \frac{1}{2} \log  \langle  \hat n_{\text{exc}} \rangle 
 \quad \quad \quad \text{for} \quad \quad \langle  \hat n_{\text{exc}}\rangle \gg 1 \ .
\end{align}
 \end{subequations}
Equations \eqref{eq:EE} and \eqref{eq:EEandnexc} 
are valid both in and out of equilibrium.  They highlight that the reduced state of subsystems $A$ and $B$ is pure (i.e., $\det G_A = 1/4$) only in fully polarized states, in which no collective spin excitations are present in the system, i.e., $\langle  \hat n_{\text{exc}} \rangle=0$.

The above equations may be seen as a direct relation between bipartite entanglement entropy and collective spin squeezing, usually quantified by the minimal transverse variance of collective spin fluctuations  \cite{HiroshimaSqueezing,WinelandSqueezing,SpinSqueezingReview}
\beq
\xi^2  \equiv \frac{\Min_{\abs{\mathbf{u}}=1,\mathbf{u}\perp \mathbf{Z}} \Big\langle \Big(\mathbf{u}\cdot \bold {\hat S}\Big)^2 \Big\rangle}{N/4} \ .
\eeq
This squeezing parameter $\xi$ is equal to $1$ for fully polarized states, while $\xi<1$ for squeezed states.
The number of collective excitations $\langle  \hat n_{\text{exc}} \rangle$ can be taken as a measure of collective spin squeezing  for general Gaussian states: from Eqs.~\eqref{eq:GScorrelations} and \eqref{eq:GSexcitations} one derives $\xi$ as
\beq
\xi^2 = e^{-2\lvert \gamma \rvert} = 1+2 \langle  \hat n_{\text{exc}} \rangle - 2 \sqrt{ \langle  \hat n_{\text{exc}}\rangle(1+ \langle  \hat n_{\text{exc}}\rangle)} \ .
\eeq
This relation shows that the amount of collective excitations $\langle  \hat n_{\text{exc}} \rangle$  increases from $0$ to $\sim N$, as
 the collective spin state is squeezed from a fully polarized configuration with $\xi=1$ 
 towards massively squeezed configurations with $\xi \sim 1/\sqrt{N}$. 

Besides the number of collective  excitations $\langle  \hat n_{\text{exc}} \rangle$ and the spin-squeezing parameter $\xi$, it is also possible to characterize entanglement entropy via yet another significant quantity, i.e., the \textit{effective temperature} of the two subsystems. In fact, the reduced density matrices may be written as $ \ro_{A,B} = Z_{A,B}^{-1} \exp (-\beta_{\text{eff}} \hat H_{A,B})$, where the state-dependent quadratic operators $\hat H_{A,B}$ are usually termed \emph{modular} or \emph{entanglement Hamiltonian}. It is straightforward to derive a relation between the effective dimensionless inverse temperature and the other quantities, e.g.,
\beq
\beta_{\text{eff}} = 2 \arctanh \bigg(  \frac{1}{\sqrt{1+4 f_A f_B \langle  \hat n_{\text{exc}}\rangle }}  \bigg).
\eeq
This equation makes it explicitly clear that the growth of $\langle  \hat n_{\text{exc}}(t)\rangle$ --- which comes via collective spin squeezing --- is responsible for ``heating  up'' the two subsystems, i.e., for raising their effective temperature, thus continuously accumulating \emph{entanglement} entropy.

\section{Nonequilibrium growth of collective excitations}

\label{sec:neqgrowth}

The polynomial growth in time of $\langle \hat n_{\text{exc}}(t) \rangle$ after a typical quench is tightly related to the well-known fact that in classical systems with a single degree of freedom, trajectories originating from nearby initial conditions separate linearly in time. In fact, as discussed above, the dynamics of a fully-connected spin system of the form in Eq. (1) of the Letter
can be rigorously described in terms 
of quantum fluctuations $G^{QQ}(t),G^{QP}(t),G^{PP}(t)$ around the classical trajectory of a single spin on the sphere, parameterized by the angles $\theta(t)$, $\phi(t)$. 
The time-dependent quadratic Hamiltonian with coefficients $\widetilde{h}^{(2)} $ 
governs the quantum evolution of $(Q, P)$ in Eq.~\eqref{eq:eom} 
and 
describes the classical evolution of small displacements around the trajectory in the linear approximation,  as well. 
Linear-in-time growth of these displacements is encoded by $U(t) \sim t$ in Eq.~\eqref{eq:U(t)} and hence by $G(t) \sim t^2$ in Eq.~\eqref{eq:G(t)}, which directly leads to $S_A(t) \sim \log t$ by Eqs.~\eqref{eq:EE} and \eqref{eq:EEandnexc}, as claimed.


We now make the above argument precise. A time-independent system with a single degree of freedom is integrable, due to the conservation of energy, and hence canonical action-angle variables $(\mathcal{A}, \varphi)$ can be introduced, where the action $\mathcal{A}$ is a constant of motion related to the area enclosed by a trajectory in phase space, and the angle $\varphi$ sweeps periodically the range $[0,2\pi]$ along the trajectory. In these variables, the (classical and quantum) evolution of the system is similar to that of a free particle,
\beq
\left\{
\begin{split}
\dot{\hat{\mathcal{A}}} &= 0 \\
\dot{\hat{\varphi}} &= \partial_{\mathcal{A}} \hat{H} \equiv \omega(\hat{\mathcal{A}}) ,
\end{split}
\right. 
\eeq 
with the solution
\beq
\left\{
\begin{split}
\hat{\mathcal{A}}(t) &= \hat{\mathcal{A}}(0) \\
\hat{\varphi}(t) &= \hat{\varphi}(0) + \omega\big(\hat{\mathcal{A}}(0)\big) t  .
\end{split}
\right. 
\eeq 
For a given classical trajectory characterized by a value of the action $\mathcal{A}_{\text{cl}}$, the evolution of quantum fluctuations  around it,
\beq
\begin{split}
\delta \hat{\mathcal{A}}(t) &\equiv \hat{\mathcal{A}}(t)- \mathcal{A}_{\text{cl}} ,\\
\delta \hat{\varphi}(t) &\equiv \hat{\varphi}(t)- \varphi_{\text{cl}}(t),
\end{split}
\eeq
 is described by
\beq
\left\{
\begin{split}
\delta \hat{\mathcal{A}}(t) &= \delta \hat{\mathcal{A}}(0) \\
\delta \hat{\varphi}(t) &=\delta \hat{\varphi}(0) +  \Big[ \omega\big(\hat{\mathcal{A}}(0)\big) - \omega\big(\mathcal{A}_{\text{cl}}\big) \Big] \, t =
   \delta \hat{\varphi}(0) +   \partial_{\mathcal{A}}\omega\big\rvert_{\mathcal{A}_{\text{cl}}}  \, \delta \hat{\mathcal{A}}(0) \, t
   + \mathcal{O} \bigg( \frac{t}{N}    \bigg) .
\end{split}
\right.
\eeq
The error term follows from the fact that the variables $(\hat{\mathcal{A}},\hat{\varphi})$ parameterize the rescaled collective spin $\mathbf{\hat{S}}/N$, 
and hence their ground state quantum fluctuations are subextensive, i.e., $\big(\delta \hat{\mathcal{A}}(0)\big)^2 \sim 1/N$. 
The time-dependence of correlations can then  be derived from the above solution,
\beq
\label{eq:GAphi}
\left\{
\begin{split}
G^{\mathcal{A}\mathcal{A}}(t) \equiv & \big\langle \delta \hat{\mathcal{A}}(t) \delta \hat{\mathcal{A}}(t) \big\rangle
= 
   G^{\mathcal{A}\mathcal{A}}(0)  \\
G^{\mathcal{A}\varphi}(t) \equiv & \frac{1}{2}\big\langle \delta \hat{\mathcal{A}}(t) \delta \hat{\varphi}(t) + \delta \hat{\varphi}(t) \delta \hat{\mathcal{A}}(t) \big\rangle  
=
G^{\mathcal{A}\varphi}(0) + \partial_{\mathcal{A}}\omega\big\rvert_{\mathcal{A}_{\text{cl}}}  G^{\mathcal{A}\mathcal{A}}(0) \, t \\ 
G^{\varphi\varphi}(t) \equiv & \big\langle \delta \hat{\varphi}(t) \delta \hat{\varphi}(t)  \big\rangle   
= 
G^{\varphi\varphi}(0) + 2 \partial_{\mathcal{A}}\omega\big\rvert_{\mathcal{A}_{\text{cl}}} \, G^{\mathcal{A}\varphi}(0) \, t 
+   \partial_{\mathcal{A}}\omega\big\rvert^2_{\mathcal{A}_{\text{cl}}}  \,  G^{\mathcal{A}\mathcal{A}}(0) \, t^2 
+  \mathcal{O} \bigg( \frac{t^2}{N}    \bigg) . \\
\end{split}
\right.
\eeq
This $t^2$-growth of quantum fluctuations is analogous to the spreading of wavepackets of free quantum particles.
Both $(\hat Q, \hat P)$ and $(\delta\hat{\mathcal{A}},\delta\hat{\varphi})$ describe quantum fluctuations of the collective spin, hence they must be related via a linear canonical transformation which depends on the instantaneous classical configuration, $(\theta,\phi)$ or $(\mathcal{A}_{\text{cl}},\varphi_{\text{cl}})$. For a general closed trajectory, the latter varies periodically in time, with a period $T_{\text{cl}}=2\pi/\omega(\mathcal{A}_{\text{cl}})$. Thus, the correlations $G^{QQ}(t)$, $G^{QP}(t)$ and $G^{PP}(t)$ are obtained from those in Eq.~\eqref{eq:GAphi} by a time-periodic linear transformation.

We have proved that the time-dependence of correlations $G^{QQ}(t)$, $G^{QP}(t)$ and $G^{PP}(t)$ generically shows a $t^2$-growth after a quench with a periodic modulation superimposed, the periodicity being that  of the underlying classical trajectory, i.e., $T_{\text{cl}}$. From Eqs.~\eqref{eq:GAphi}, we see that corrections to this behavior manifest over the Eherenfest time scale $t_{ \text{Ehr}} \sim \sqrt{N}$, which diverges in the thermodynamic limit. We emphasize that all the quantities above for a given system can in principle be computed analytically, as the system is Liouville-integrable \cite{landau}.

%
%


A remarkable exception to the behavior described above is represented by isolated trajectories in phase space known as \textit{separatrices}, which traverse unstable fixed points and divide the sphere into topologically distinct disconnected regions, and which correspond to a singularity of the action-angle variables. These trajectories have a divergent period 
and are  related to the so-called \emph{mean-field dynamical criticality} \cite{SciollaBiroliMF}.
For quenches to such dynamical critical points, the growth of the number of collective excitations is exponential in time rather than polynomial, due to the exponential separation of classical trajectories originating from points near a separatrix. The rate of such an exponential separation is determined by the positive eigenvalue $\lambda$ of the linearized flow around the unstable fixed point [see Fig.~2(b) of the Letter, for an illustration]. In fact, exponential growth of quantum fluctuations is encoded by $U(t) \sim \exp (\lambda t)$ in Eq.~\eqref{eq:U(t)} and hence by $G(t) \sim \exp (2 \lambda t)$ in Eq.~\eqref{eq:G(t)}, which directly leads to $S_A(t) \sim \lambda t$ by Eqs.~\eqref{eq:EE} and \eqref{eq:EEandnexc}, as claimed. The rigorous proof of this is essentially analogous to that presented in the previous above for generic quenches. This effect is similar to what is discussed in Ref.~\onlinecite{zurek1995quantum} in the context of open quantum systems. 

\section{Long-time saturation of entanglement entropy}
\label{sec:saturation}
As remarked above, 
our approach to the non-equilibrium dynamics is adequate only before the \emph{Ehrenfest time scale}  
in correspondence of which the description becomes inaccurate, i.e., $\big\langle \hat n_{\text{exc}}(t=t_{ \text{Ehr}})
\big\rangle \sim N$: in fact, at such time scale the quantum fluctuations of the collective spin become comparable with the magnitude of the collective spin itself.
This time scale diverges with the system size in a way which depends on the nature of the underlying semiclassical trajectories, i.e., 
$t_{ \text{Ehr}} \sim \sqrt{N}$ for generic quenches and $t_{ \text{Ehr}} \sim \log N$ for quenches to dynamical critical points, 
 and, accordingly, it sets the limit of validity of semiclassical analyses  \cite{schubert2012how}.
%
At this time scale, the number of excitations reaches its maximal value, implying through Eq.~(\ref{eq:big_S})
 a saturation value of $S_A$ proportional to the logarithm of the number of spins in subsystem $A$,
\beq
\label{eq:sat}
 S_A  \sim  \frac{1}{2} \log  N_A \ .
 \eeq
This is actually related to the usual volume-law scaling of entanglement out of equilibrium. 
In fact, the stationary states after a quantum quench explore all the allowed Hilbert space, and their entanglement is upper-bounded by $S_A \le \log \left (\text{dim} \mathcal H_A\right )$.
For generic many-body systems, the dimension of $\mathcal H_A$ is exponentially large with the volume of the subsystem [e.g. $\text{dim}(\mathcal{H}_A) = 2^{N_A}$ for 
spins-$1/2$
], causing volume-law scaling.
In collective models under consideration here, however, the conservation of the collective spin magnitude $\lvert \hat{\mathbf{S}} \rvert^2$ reduces the dimension of the allowed Hilbert space to $\text {dim} (\mathcal{H}_A) = N_A + 1$.
For an illustration with numerical results on the long-range quantum Ising model, see Fig. 3 of the Letter.



\section{Spatially-decaying interactions}

\label{sec_decaying}


\label{eq:alphaneq0} 
We deal with a quantum spin-$1/2$ system on a $d$-dimensional cubic lattice of size $L$ governed by a Hamiltonian of the form
\beq
\label{eq:Hlongrange}
\hat H= - \frac 1{\mathcal N_{\alpha,N}} \sum_{\mu, \nu=x,y,z} J_{\mu\nu} \sum_{i\neq j}^N \frac{\hat s_i^{\mu}\hat s_j^{\nu}}{\lvert\mathbf{r}_i-\mathbf{r}_j\rvert^{\alpha}}
- \sum_{\mu=x,y,z} h_{\mu} \sum_{i}^N \hat s_{i}^{\mu} \ , 
\eeq
where the exponent $\alpha\ge0$ characterizes the algebraic decay of spin-spin interactions. The distance $\lvert\mathbf{r}_i-\mathbf{r}_j\rvert$ between two sites on the periodic  lattice  can be taken to be $\sqrt{\sum_{\mu=1}^d \big[ \Min(|r_i^\mu - r_j^\mu|, L-|r_i^\mu - r_j^\mu |)\big]^2}$. A Ka\v{c} rescaling factor  $1/\mathcal{N}_{\alpha,N}$ with $\mathcal{N}_{\alpha,N}=\sum_{ i \ne j } \lvert\mathbf{r}_i-\mathbf{r}_j\rvert^{-\alpha} / N$ replaces the $1/N$ factor in Eq.~(1) of the Letter, ensuring the extensivity of the Hamiltonian for $\alpha\le d$ \cite{KacNormalization}. The fully-connected limit is recovered by letting $\alpha\to0$.

\subsection{Nonequilibrium spin-wave expansion}

We now refine the two-boson formalism above to make it suitable for many-body problems. We expand the individual spins 
around the  instantaneous direction $\mathbf{Z} \parallel \langle \mathbf{S}(t) \rangle$ of the collective spin 
via Holstein-Primakoff transformations,\cite{LeroseShort,LeroseLong} 
\beq
\label{eq:H-Pindividual}
\bold {\hat s}_j \; \simeq \; \sqrt{s} \, \mathbf{X} \, \hat q_j \, + \sqrt{s} \, \mathbf{Y} \, \hat p_j   + \mathbf{Z} \, \bigg( s-\frac{\hat q_j^2+\hat p_j^2-1}{2} \bigg)
\eeq
\noindent where $s=1/2$ and the time-dependent rotated frame $(\mathbf{X},\mathbf{Y},\mathbf{Z})$ parameterized by spherical angles $\theta$, $\phi$ was introduced in Eq.~\eqref{eq:newbasis}. 
The rotating-frame Hamiltonian 
$
\widetilde{\hat H}(t) = \hat H - \boldsymbol{\omega}(t) \cdot \bold {\hat S}
$
[cf. Eq.~\eqref{eq:Hrotatingframe}] 
can then be expanded through Eq.~\eqref{eq:H-Pindividual}
in terms of the \textit{spin-wave} variables $\tilde{q}_\mathbf{k}= L^{-d/2} \sum_j e^{-i\mathbf{k}\cdot \mathbf{r}_j} \hat q_j$ and $\tilde{p}_\mathbf{k}= L^{-d/2} \sum_j e^{-i\mathbf{k}\cdot \mathbf{r}_j} \hat p_j$ at all possible momenta $\mathbf{k}$. One obtains
%
%
\beq
\label{eq:Hsw}
\widetilde{\hat H}(t) =  
 \; \widetilde{\hat H}_{0}(t) -  \sum_{\mathbf{k}\neq 0}  \widetilde{f}_{\alpha,\mathbf{k}} \bigg[ 
     J_{qq}(\theta, \phi) \; \frac{\tilde{q}_\mathbf{k}\tilde{q}_{-\mathbf{k}}}{2} +
      J_{pp}(\theta, \phi) \; \frac{\tilde{p}_\mathbf{k} \tilde{p}_{-\mathbf{k}}}{2} 
      +
        J_{qp}(\theta, \phi) \; \frac{\tilde{q}_\mathbf{k} \tilde{p}_{-\mathbf{k}}+\tilde{p}_\mathbf{k} \tilde{q}_{-\mathbf{k}} }{2}
       \bigg] + \mathcal{O} \bigg( \frac{1}{\sqrt{N}} \bigg),
\eeq
where $J$'s are coefficients depending on the angles $\theta(t)$ and $\phi(t)$ and
\beq
\label{eq:Btildek}
\widetilde{f}_{\alpha,\mathbf{k}} = \frac{1}{\mathcal{N}_{\alpha,N}} \sum_{j (\ne i)} \frac{e^{-i\mathbf{k}\cdot (\mathbf{r}_j-\mathbf{r}_i)}}{\lvert\mathbf{r}_j-\mathbf{r}_i\rvert^{\alpha}} ,
\eeq
and the collective-mode Hamiltonian
\beq
\label{eq:Hsw0}
\widetilde{\hat H}_0(t) =  
 \; N \, \mathcal E_{\text{cl}}(\theta, \phi) 
\; - 
\bigg[   J_{qq}(\theta, \phi)  \frac{\hat Q^2}{2}
    + J_{pp} (\theta, \phi) \frac{\hat P^2}{2}
    + J_{qp}(\theta, \phi)  \frac{\hat Q \hat P+\hat P \hat Q}{2}
    \bigg] 
    +
\widetilde{h}^{(2)}_{\text{sw}}
\,  \sum_{\mathbf{k}} \hat n_{\mathbf{k}} 
    + \mathcal{O} \bigg( \frac{1}{\sqrt{N}} \bigg) \, 
\eeq
accounts for the infinite-range part 
$\widetilde{f}_{\alpha,\mathbf{0}} \, \delta_{\mathbf{k},\mathbf{0}} \equiv \delta_{\mathbf{k},\mathbf{0}}$ of the interaction $\widetilde{f}_{\alpha,\mathbf{k}}$.
The collective-mode Hamiltonian $\widetilde{H}_{0}(t)$ describes the non-trivial dynamics of collective spin fluctuations $\hat Q\equiv \tilde{q}_\mathbf{0}$ and $\hat P \equiv \tilde{p}_\mathbf{0}$,  
but conserves the bosonic occupation numbers of all the other spin-wave modes at finite wavelength, 
%
\beq
\label{eq:boso_k}
\hat n_{\mathbf{k}\ne\mathbf{0}} \equiv \frac{\tilde{q}_\mathbf{k} \tilde{q}_{-\mathbf{k}}+\tilde{p}_\mathbf{k} \tilde{p}_{-\mathbf{k}}-1}{2}\ .
\eeq
as  $\Big[\hat n_\mathbf{k}, \widetilde{\hat H}_0 \Big] = 0 $ for all $\mathbf{k}\ne0$ (note that this is rigorously true to all orders in the Holstein-Primakoff expansion).
In 
the above equations, it is assumed that the motion of the angles $\theta(t)$ and $\phi(t)$ is fixed in such a way that linear terms in the collective quantum fluctuations $Q\equiv \tilde{q}_\mathbf{0}$ and $P \equiv \tilde{p}_\mathbf{0}$ vanish, which is equivalent to the self-consistency requirement ${\langle \hat S^X(t) \rangle \equiv \langle \hat S^Y(t) \rangle \equiv 0}$. In this case, the dynamical generation of spin waves can modify the classical trajectory of the collective spin \eqref{eq:motion_angles0}, due to the feedback from quantum fluctuations \cite{LeroseShort,LeroseLong}.

\subsection{Bounds for $\tilde{f}_{\alpha,\mathbf{k}}$}

\begin{figure}[H]
\centering
\includegraphics[width=0.7\textwidth]{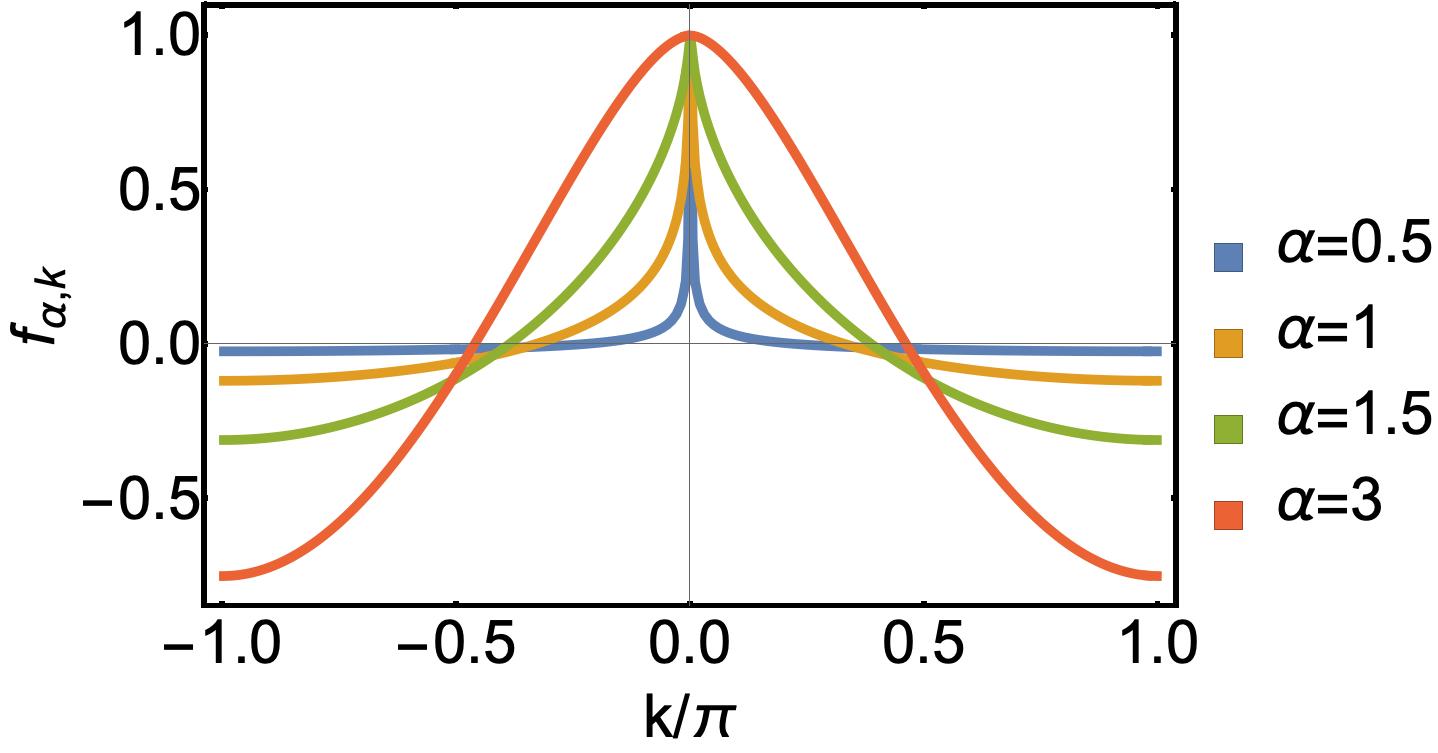} \\
\includegraphics[width=0.49\textwidth]{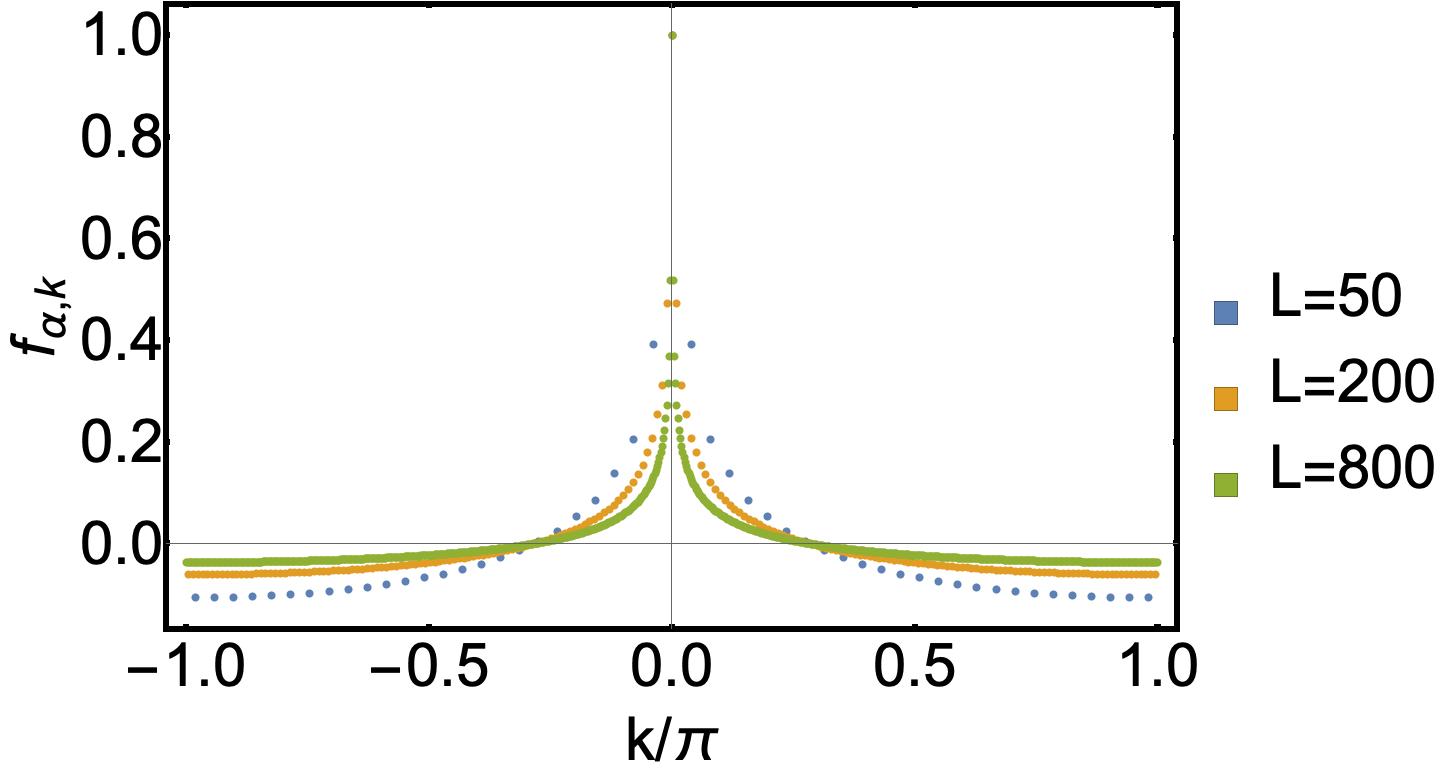} 
\includegraphics[width=0.49\textwidth]{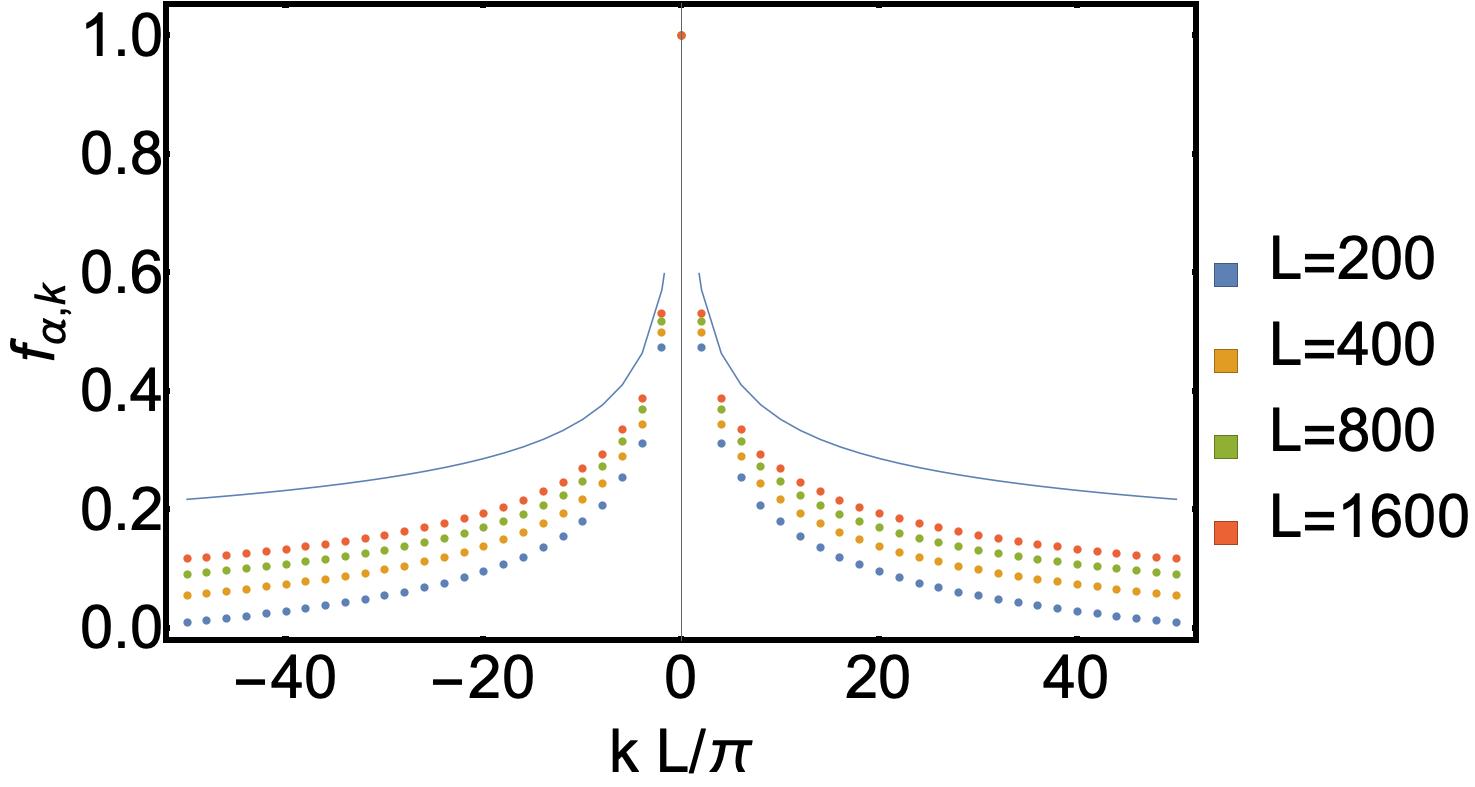} \\
\caption{
{Plots of the function $\widetilde{f}_{\alpha,k}$ for $d=1$.
Top panel:  $\widetilde{f}_{\alpha,k}$ is shown for several values of $\alpha$, for $N=L=400$.
One  recognizes a function squeezed towards $k=0$ ($0\le \alpha \le 1$), a finite function with a cusp behavior for small $k$ ($1<\alpha<2$), and a cosine-like function ($\alpha\gg2$).
Bottom left panel: $\widetilde{f}_{\alpha,k}$ is shown for $\alpha=0.7$ and increasing values of $N=L$.
A qualitatively similar behavior occurs for $0\le \alpha \le 1$.
Squeezing towards a delta function as $L\to\infty$ occurs with a speed $N^{-(1-\alpha)}$ for $\alpha<1$ and $1/\log N$ for $\alpha=1$.
Bottom right panel: a "zoom" of the plot in the bottom left panel is shown, for larger  values of $N=L$.
The \emph{rescaled} function in the vicinity of $k=0$  converges to a finite limiting curve as $L\to\infty$.
This discrete structure approaches a continuum as $\alpha \nearrow 1$.
The blue solid  line illustrates the bound obtained in Eq. \eqref{eq_bounds}.}
}
\label{fig_ftilde}
\end{figure}

In view of the above discussion,
the corrections due to weak spatial decay of interactions  $\alpha>0$ to entanglement dynamics associated with collective spin excitations,  discussed in Sec. \ref{sec:neqgrowth}
 above, are encoded in the analytic structure of the couplings $\tilde{f}_{\alpha,\mathbf{k}}$ in Eq.~\eqref{eq:Hsw},
 which are vanishing in the infinite-range model with $\alpha=0$. 
Assuming long-range interactions with $\alpha < d$, we can safely approximate sums with integrals in their defining expression \eqref{eq:Btildek}, which captures the leading order exactly. Hence, we can switch to spherical coordinates and integrate over all the angles, obtaining
\beq
\widetilde{f}_{\alpha,\mathbf{k}\ne\mathbf{0}} \thicksim \frac{1}{L^{d-\alpha}} \int_1^{L} d\rho \, \rho^{d-1-\alpha} \, 
\frac{\mathcal{J}_{d/2-1} (\abs{\mathbf{k}}\rho)}{(\abs{\mathbf{k}} \rho)^{d/2-1}},
\eeq
where $\mathcal{J}_\nu(x)$ is the standard Bessel function of order $\nu$.
While for small $\rho$ the integral is never singular (due to the assumption of long-range interactions, $\alpha<d$),
for large $\rho$ the integrand is asymptotically oscillating with period $2\pi/\abs{\mathbf{k}}$ and amplitude decaying as $\rho^{(d-1)/2-\alpha}$, which yields convergence only for $\alpha>(d-1)/2$.
In this case, by rescaling $\abs{\mathbf{k}}\rho = \eta$ to obtain a dimensionless integrand and denoting by $F(\eta)$ its primitive which satisfies $F(\infty)=0$ and is uniformly bounded, one obtains the asymptotic estimate
\beq
\widetilde{f}_{\alpha,\mathbf{k}\ne\mathbf{0}} \; \underset{L\to\infty}{\thicksim} \; - \frac{F(\abs{\mathbf{k}})}{(\abs{\mathbf{k}}L)^{d-\alpha}} .
\eeq
In the limiting case $\alpha=d$, one can similarly compute
\beq
\widetilde{f}_{\alpha,\mathbf{k}\ne\mathbf{0}} \; \underset{L\to\infty}{\thicksim} \; - 
\frac{\log \abs{\mathbf{k}} + \tilde{F}(\abs{\mathbf{k}})}{\log  L} ,
\eeq
where $\tilde{F}$ is the non-singular (bounded) part of the primitive $F$.
On the other hand, for $\alpha\le (d-1)/2$, by isolating the purely oscillatory terms and repeatedly integrating by parts, we can obtain an expansion of the primitive of the form
\beq
\widetilde{f}_{\alpha,\mathbf{k}\ne\mathbf{0}} \thicksim \frac{1}{(\abs{\mathbf{k}}L)^{d-\alpha}} 
\bigg( 
c_1
(\abs{\mathbf{k}}L)^{\frac{d-1}{2}-\alpha}
+ c_2 
(\abs{\mathbf{k}}L)^{\frac{d-1}{2}-\alpha-1}
+\dots
\bigg)
\eeq
with $c_{1,2,\dots}$ numerical constants.
To sum up,
 the following estimates have been established, 
\beq
\label{eq_bounds}
\big\lvert \widetilde{f}_{\alpha,\mathbf{k}\ne\mathbf{0}} \big\rvert \; \le \;  
\left\{
\begin{split}
\text{const } & \times \frac{1}{(\abs{\mathbf{k}} L)^{\frac{d+1}{2}}}  &  \text{for }&  \alpha \le \frac{d-1}{2} , \\
\text{const } & \times \frac{1}{(\abs{\mathbf{k}} L)^{d-\alpha}} \quad & \text{for }& \frac{d-1}{2} < \alpha <d, \\
\text{const } & \times \frac{\big \lvert \log \abs{\mathbf{k}} + \tilde{F}(\abs{\mathbf{k}})  \big\rvert}{\log  L} \quad & \text{for }& \alpha =d .
\end{split}
\right. 
\eeq
{See Fig. \ref{fig_ftilde} for an illustration of the behavior of the function $\widetilde{f}_{\alpha,k} $ in $d=1$.} \\

\subsection{Dynamics of the spin-wave population and entanglement}

{
We can now analyze the full contribution of the spin waves to the entanglement entropy growth, which improves on the spin-squeezing-induced entanglement growth expressed by Eq. (\ref{eq:EE}).
To this end, we first discuss the evolution of the population of the spin waves after a quench.
}

{
Equation \eqref{eq:Hsw} shows that, within the linear spin-wave analysis, the system is equivalent to a set of \emph{periodically driven quantum harmonic oscillators}, labelled by the quasimomentum $\mathbf{k}$. The classical evolution of the collective spin, described by the periodic dynamics of the angles $\theta(t)$, $\phi(t)$ with a frequency $\omega_{\text{cl}} \equiv 2\pi/T_{\text{cl}}$, acts as a drive on these bosonic modes  described by the variables $(\tilde{q}_{\mathbf{k}} (t),\tilde{p}_{\mathbf{k}} (t))$. 
The driving frequency $\omega_{\text{cl}}$  is common to all $\mathbf{k}$'s and depends only on the quench, while the driving amplitude depends both on the quench and on $\mathbf{k}$ via the coupling strength $\widetilde{f}_{\alpha,\mathbf{k}}$.
As a consequence of the bounds in Eqs. (\ref{eq_bounds}), the driving amplitude is vanishingly small in the thermodynamic limit for all fixed $\mathbf{k}$'s when $\alpha < d$, which implies that the excitation of all finite-wavelength modes is vanishingly weak, so that their effects are generically negligible for large $N$ --- and in any case delayed to the divergent time scale $T_{\text{sw}}\sim N^{\beta/d}$ with $\beta\equiv\Min\big(d-\alpha,(d+1)/2\big)$. 
(Note that this a posteriori fully validates the Holstein-Primakoff expansion and the absence of corrections to the collective spin trajectory for $\alpha<d$, see also Refs. \onlinecite{LeroseShort,LeroseLong,LeroseKapitza}.) However, long-wavelength modes with $\abs{\mathbf{k}} \propto 1/L$ are driven with a finite intensity. Therefore, in the presence of long-range interactions with $\alpha < d$, the description of the system dynamics effectively reduces to a \emph{discrete} set of driven quantum oscillators, corresponding to spin fluctuations with $k_\mu = 0, \pm 2\pi/L, \pm 4\pi/L, \dots, \pm 2\pi n^*/L$, for $\mu=1,\dots,d$, where the cutoff $n^* \gg 1$ can be taken independent of $L$, see Fig. \ref{fig_ftilde}, bottom right panel.
This is in contrast with systems with shorter-range interactions (i.e., with $\alpha>d$), in which a \emph{continuum} of traveling quasiparticles generate linear growth of entanglement entropy through the standard Calabrese-Cardy mechanism \cite{CC_QuasiparticlePicture}.
}

{
For a large set of initial conditions, spin waves are \emph{non-resonantly driven}, and consequently their population remains bounded in time. In this case, the spin-squeezing contribution captures the leading behavior of entanglement entropy, and the contribution of spin waves represents a finite correction to the dominant logarithmic growth characteristic of integrable semiclassical dynamics.
However, particular quenches 
typically near mean-field dynamical critical points, may give rise to a \emph{resonant excitation} of long-wavelength spin waves, leading to an exponentially growing population thereof. 
This occurrence is triggered by the well-known mechanism of ``parametric resonance'' in driven oscillators
~\cite{landau}.
This effect is a hallmark of semiclassical chaos induced by the finiteness of the interaction range and is associated with a linear increase of entanglement entropy in time (see below).
}

{
A \emph{stability analysis} of the spin-wave excitations allows one to predict the nature of entanglement growth (logarithmic or linear) for any given  quench.
It  can be performed as follows.
For any long-range spin Hamiltonian, one can perform a nonequilibrium spin-wave expansion as explained in Sec. \ref{sec_decaying}A above, and compute the {time-evolution} 
\beq
\begin{pmatrix}
\tilde{q}_{\mathbf{k}}(t_0+T_{\text{cl}}) \\
\tilde{p}_{\mathbf{k}}(t_0+T_{\text{cl}})
\end{pmatrix} 
= U_{\mathbf{k}}(T_{\text{cl}}) \cdot
\begin{pmatrix}
\tilde{q}_{\mathbf{k}}(t_0) \\
\tilde{p}_{\mathbf{k}}(t_0)
\end{pmatrix} 
\eeq
of each spin-wave mode {over one classical period} $T_{\text{cl}}$. 
Hence, one obtains the  eigenvalues $e^{\pm \lambda_{\mathbf{k}} T_{\text{cl}}}$ of the $2 \times 2$ matrix $U_{\mathbf{k}}(T_{\text{cl}})$.
The number $\lambda_{\mathbf{k}}$, known as the \emph{Floquet quasi-frequency} (see, e.g., Refs. \onlinecite{landau,BlanesMagnusReview}) of the driven oscillator, determines the resonance condition of the driven oscillator. 
If $\lambda_{\mathbf{k}}=i\omega_{\mathbf{k}}$ is \emph{purely imaginary}, then the mode is stable and its amplitude remains bounded in time, oscillating at a frequency $\abs{\omega_{\mathbf{k}}}$. 
On the contrary, if $\lambda_{\mathbf{k}}$ is \emph{real}, the mode in unstable and its amplitude grows exponentially fast in time with a rate $\abs{\lambda_{\mathbf{k}}}$.
Isolated resonances may in principle occur for particular trajectories. 
It seems to be typically the case (see the examples below) that quenches \emph{near} dynamical criticality give rise to resonant excitation of spin waves. 
In other words, the classical separatrix of the mean-field dynamics for $\alpha=0$ broadens to a \emph{finite} layer of instability (chaoticity) for $\alpha>0$.
}


{
The above discussion allows us to understand the full spin-wave 
time-dependent entanglement entropy, using known mathematical results for quadratic bosons \cite{BianchiModakRigolEntanglementBosons}.
Even for $\alpha>0$, the entanglement between two subsets of quantum spins is encoded in the entanglement between their respective bosonic fluctuations.
For a general system of quadratic bosons, $S_A(t)$ can be computed with standard techniques --- see, e.g., Refs. [\onlinecite{BarthelEEquadratic,gaussianQI,BianchiModakRigolEntanglementBosons}].
We briefly recall how this computation can be performed for translation-invariant Hamiltonians such as Eq. \eqref{eq:Hsw}.}
{The time evolution is diagonal in Fourier space, and one integrates $N$ decoupled pairs of equations of motion for $(\tilde{q}_{\mathbf{k}} (t),\tilde{p}_{\mathbf{k}} (t))$ to obtain the time-evolved Gaussian state of each Fourier mode, described by the correlations $(\tilde{G}^{qq}_{\mathbf{k}}(t),\tilde{G}^{qp}_{\mathbf{k}}(t),\tilde{G}^{pp}_{\mathbf{k}}(t))$ defined by
\beq
\tilde{G}^{\alpha\beta}_{\mathbf{k}}(t) = 
\frac{1}{2} \Big\langle  
\tilde{\alpha}_{\mathbf{k}} (t) \tilde{\beta}_{-\mathbf{k}} (t) + \tilde{\beta}_{\mathbf{k}} (t) \tilde{\alpha}_{-\mathbf{k}} (t)
\Big\rangle
\eeq
for $\alpha,\beta=q,p$.
The time-evolved operators $\tilde{q}_{\mathbf{k}} (t)$, $\tilde{p}_{\mathbf{k}} (t)$ are linearly related to those at time $t=0$ via the solution to the Heisenberg equations of motion $\dot {\tilde{\alpha}}_{\mathbf{k}} (t) = -i \big[ \tilde{\alpha}_{\mathbf{k}} (t), \widetilde{\hat H}(t) \big]$ with the generator $\widetilde{\hat H}(t)$ in Eq. \eqref{eq:Hsw}.}
{Within the linear spin-wave analysis, the state of a subsystem composed of $M < N$ spins contained in a region $A$ of the lattice is a Gaussian bosonic state determined by the instantaneous correlations 
\beq
\bigg\{ G^{\alpha\beta}_{ij}(t) =  \big\langle  
\alpha_i (t) \beta_j (t) + \beta_i (t) \alpha_j (t)
\big\rangle \bigg\}_{\substack{i,j\in A \\ \alpha,\beta=q,p}}
\eeq
 between them,
 which can be expressed in terms of $\tilde{G}^{\alpha\beta}_{\mathbf{k}}(t)$ via Fourier antitransform:
 \beq
 \label{eq:corre_matrix}
 G^{\alpha\beta}_{ij}(t) =  \frac{2}{N} \sum_\mathbf{k} e^{i\mathbf{k}(\mathbf{r}_i-\mathbf{r}_j)} \tilde{G}^{\alpha\beta}_{\mathbf{k}}(t) \equiv G^{\alpha\beta}_{\mathbf{r}}(t) \Big\rvert_{\mathbf{r}=\mathbf{r}_i-\mathbf{r}_j} \; .
 \eeq
This set of correlations for $i,j \in A$, collected in a $2M \times 2M$ matrix $G_A$, uniquely identifies  the reduced density matrix $\hat \rho_A (t)$. 
}
{
The von Neumann entropy of this Gaussian bosonic state 
can be
 computed as \cite{BarthelEEquadratic,gaussianQI,BianchiModakRigolEntanglementBosons}
\begin{equation}
\label{eq:ee_sw}
S_A = \sum_{i=1}^M\, S(\nu_i) \ , \quad \text{with} \quad 
S(\nu_i) = \frac{\nu_i+1}2\, \ln \frac{\nu_i+1}2 - \frac{\nu_i-1}2\, \ln \frac{\nu_i-1}2 \ ,
\end{equation}
where $\nu_i$ are the so-called \emph{symplectic eigenvalues} of the correlation matrix, defined as follows. From the correlation function in Eq. (\ref{eq:corre_matrix}), one defines the matrix $J=-G\, \Omega$, where $\Omega$ is the $2N\times 2N$ symplectic unit. 
The matrix $[iJ]_A$ restricted to $A$ can be shown to have pairs of opposite real eigenvalues $\pm \nu_i$ (with $\nu_i > 1$ as follows from the Heisenberg relations). 
The numbers $\{ \nu_i \}_{i=1,\dots,M}$ are referred to as the symplectic eigenvalues of $G_A$, and determine the entropy via Eq. \eqref{eq:ee_sw}. 
}

{
For long-range interactions with $0<\alpha<d$, the growth of $S_A(t)$ turns out to be determined by the stability of the discrete set of long-wavelength excitations, expressed by the Floquet quasi-frequencies $\lambda_{\mathbf{k}}$ with $\abs{\mathbf{k}} \propto 1/L$: see the above discussion, Fig. \ref{fig_ftilde} and, e.g., Ref. \onlinecite{BianchiModakRigolEntanglementBosons}. In particular, if all of them are imaginary (i.e., all modes are stable), then $S_A(t)\sim \log t$ exhibits a slow growth dominated by the collective spin fluctuations with $k=0$. On the other hand, if some of them are real (i.e., some modes are unstable), then $S_A(t)\sim h_{KS} \, t$ exhibits a fast growth dominated by the unstable quantum fluctuations, inasmuch as  $h_{KS} = \sum_{k} \big\lvert \Re (\lambda_k)  \big\rvert$. 
(The latter quantity coincides with the well-known \emph{Kolmogorov-Sinai entropy rate}, a standard measure of chaoticity in classical dynamical systems: see, e.g.,  Ref. \onlinecite{VulpianiChaosbook}.)
In view of the above discussion of the evolution of the $k$-resolved spin-wave population after a quench, we conclude that typical quenches in a long-range interacting quantum spin-1/2 system yield a logarithmic growth of the von Neumann entanglement entropy, as argued in the main text. See below for a numerical illustration.
}

\section{Details on the numerical simulations }
\label{sec:LMG}
This Section provides  the details of our calculations for the long-range quantum Ising chain, described by the Hamiltonian \eqref{eq:Hlongrange} with $d=1$ and with non-vanishing Ising coupling $J_{xx}\equiv 2J$ and transverse field $h_z\equiv 2h$, i.e.,
\beq
\label{eq:Halpha}
\hat H = 
 -  \frac{J}{ \mathcal{N}_{\alpha,N}}  \sum_{ i < j }^N \frac{\hat \sigma_{i}^x \hat \sigma_{j}^x}{\lvert i-j\rvert^{\alpha}}
 - h \sum_{i}^N \hat \sigma_{i}^z ,
\eeq
where $\hat \sigma_i^{x,z}= 2 \hat s_i^{x,z}$ are Pauli matrices on sites $i=1,\dots,N$ of the chain, $h$ is the global transverse magnetic field and $\mathcal{N}_{\alpha,N}=\sum_{ i \ne j } \lvert i- j\rvert^{-\alpha} / N$ is the Ka\v{c} rescaling, ensuring the extensivity of the Hamiltonian for $\alpha\le d$. \\

\paragraph{The Lipkin-Meshkov-Glick model.}
We will first focus on 
its infinite-range version for $\alpha=0$, equivalent to the widely-known Lipkin-Meshkov-Glick model \cite{lipkin1965validity}. 
We  apply the general scheme and results found in Secs.~\ref{sec:holsta_prima} and \ref{app:enta_colle_exc} to show how the entanglement entropy growth is intimately related to the structure of the semiclassical trajectories. 
%

 For large values of the transverse field $\abs{h}>J$ the system is paramagnetic, with a single equilibrium configuration of the spins aligned with the field direction, and the non-equilibrium dynamics consist in a precession around it. A quantum phase transition at $h=\pm J$ separates this phase from a ferromagnetic one, with a pair of degenerate ground states with spin orientation in the $x$-$z$ plane, symmetric with respect to flipping the $x$ axis.
 The out-of-equilibrium behavior has been widely studied \cite{bapst2012quantum, russomanno2015thermalization,PhysRevB.96.104436} and,  
in the case of a quantum quench of the transverse field $h_0\to h_f$, it is characterized by the phenomenon of {dynamical phase transitions} (DPTs)  \cite{SciollaBiroliMF}. The non-equilibrium trajectories of the system may have paramagnetic or ferromagnetic character depending on the initial state and on the final transverse field $h_f$. The two families are distinguished by the time-averaged magnetization $\overline{S_x(t)}$ being vanishing or not, and are separated by a critical trajectory (\emph{separatrix}) with a diverging period, see Fig.~2 of the Letter for an illustration. 
%
The ground state entanglement entropy of the LMG model has been studied in Refs.~\onlinecite{vidal2007enta, latorre2005entanglement}, where it is found to be finite away from the quantum critical point and logarithmically divergent with the system size in correspondence of it. More recently, 
its growth in time after a quench of the transverse field has been numerically found to be consistent with a logarithmic behavior  \cite{Daley, DaleyEssler}. 
%
%
The non-equilibrium evolution governed by the Hamiltonian \eqref{eq:Halpha} with $\alpha=0$ has been studied with the dynamical approach of Sec.~\ref{sec:holsta_prima} in Refs.~\onlinecite{LeroseShort, LeroseLong, pappascrambling}.

 The expansion \eqref{eq:HQP} of the Hamiltonian in the rotating frame via Eqs.~\eqref{eq:Hrotatingframe} and \eqref{eq:H2tilde} 
in this case reads
\begin{subequations}
\begin{align}
\label{eq:Ecl}
&\widetilde{\mathcal {E}}_{\text{cl}} = - h\, \cos \theta -\frac 12 \cos \theta \dot{\phi} -\frac J2 \sin^2\theta\cos^2 \phi \ , \\ \nn \\
& \widetilde{h}^{(1)}_Q = 
  h\, \sin\theta +\frac 12 \sin \theta \, \dot{\phi} -J\cos\theta\sin\theta\cos^2\phi \ , \quad 
\widetilde{h}^{(1)}_P  = -\frac 12\, \dot{\theta} +J\sin\theta\, \sin\phi\cos\phi \ ,\\ \nn \\
& \widetilde{h}^{(2)}_{QQ} =  J \sin^2\theta \cos^2 \phi 
\ , \quad 
 \widetilde{h}^{(2)}_{PP}  = J \cos 2 \phi  
 \ , \\
& \widetilde{h}^{(2)}_{QP}   =  J \cos \theta \sin \phi\cos\phi \ , \quad 
\label{eq:hsw}
 \widetilde{h}^{(2)}_{\text{sw}}  = J \cos^2 \phi\ .
\end{align}
\end{subequations}
By setting to zero the linear terms $\widetilde{h}^{(1)}$, the classical equations of motion \cite{das06} are obtained
\begin{align}
\label{eq:motion_angles0}
\begin{dcases}
\dot \theta  =  2 J \sin\theta \cos\phi \sin\phi  \\
\dot \phi = -2h  + 2 J   \cos\theta \cos^2\phi   \ ,
    \end{dcases}
\end{align}
while the dynamical correlati ons of  collective spin fluctuations in Eq.~\eqref{eq:collective_dyn_corr} evolve according to
\begin{align}
\label{eq:motion_feedback_0}
\begin{dcases}
\dot{G}^{QQ} = 4J \cos\theta\sin\phi\cos\phi\, G^{QQ} + 4 J \cos 2 \phi\,G^{QP} \\
\dot{G}^{PP} = -4J \cos\theta\sin\phi\cos\phi\, G^{PP} - 4 J  \cos^2\phi\sin^2\theta\,G^{QP}  \\
\dot G^{QP} = -2J \cos^2\phi\sin^2\theta\,G^{QQ}  + 2J  \cos 2\phi\,G^{PP}.
\end{dcases} \ 
\end{align}
with $\theta=\theta(t)$ and $\phi=\phi(t)$ determined by Eq.~\eqref{eq:motion_angles0}. These equations are exact for $N\to \infty$, while finite-size effects become manifest at the Ehrenfest time scale $t_{ \text{Ehr}}(N)$, which depends on the nature of the semiclassical trajectory. As previously discussed, for generic quenches $t_{ \text{Ehr}}\sim \sqrt N$, while at the DPT, corresponding to the separatrix in the classical phase space, it becomes $t_{ \text{Ehr}}\sim \log N$. Equations 
\eqref{eq:motion_feedback_0}  are a set of linear time-dependent differential equations and their numerical integration with the appropriate initial conditions [given by Eq.~\eqref{eq:GScorrelations} for a general quench], determines the time-evolution of the number of collective excitations $\langle \hat n_{\text{exc}}(t)\rangle $ in Eq.~(\ref{eq:ne}). 

The analytical treatment is tested against exact-diagonalization (ED) numerical simulations. Due to the conservation of the collective spin $\lvert \bold{\hat S} \rvert^2$ in infinite-range systems, the time-evolving wavefunction is constrained within the maximal-spin Hilbert space sector $\lvert \bold{\hat S} \rvert = N/2$, whose dimension $N+1$ grows only linearly with the system size, which allows one to easily simulate the dynamics of large systems. We compute entanglement following the decomposition in Ref.~\onlinecite{latorre2005entanglement}.
We first focus on quantum quenches from a ferromagnetic ground state. For the sake of definiteness, we consider one of the two ground states of the LMG Hamiltonian (\ref{eq:Halpha}) with $h_0=0$, e.g., the one fully polarized along the positive $x$-axis,
\beq \ket{\psi_0}=\ket{\rightarrow \, \rightarrow \, \dots \rightarrow \,}.\eeq 
It corresponds to the initial conditions $\theta_0=\pi/2$, $\phi_0=0$, $G^{QP}(t=0)=0$ and $G^{QQ}(t=0)=G^{PP}(t=0)=1/2$  in Eqs.~(\ref{eq:motion_angles0}), (\ref{eq:motion_feedback_0}).  The initial state $\ket {\psi_0}$ is then evolved in the presence of a transverse field  $h=h_f$, varying above, below and at the dynamical critical  point 
$h_c = J/2$.  
As Figs.~1 and 3 of the Letter illustrate, in all the three cases the finite-size numerical result perfectly agree with the analytical result based on our general formula for $t\lesssim t_{ \text{Ehr}}$. For quenches above and below $h_c$, the entanglement entropy increases logarithmically after a transient, i.e., $S_A\sim \log t$, before saturation at $t_{ \text{Ehr}}\sim\sqrt N$, see Fig.~1 of the Letter and Fig.~\ref{fig:3} below. 
In turn, at the dynamical critical point, due to the exponential growth of the collective excitations, the entanglement entropy increases linearly in time, i.e., $S_A\sim \lambda_{h_c} t$, before saturation at $t_{ \text{Ehr}}\sim \log N$. The slope $\lambda_{h_c}$ corresponds to the instability rate of the linearized flow, i.e., the imaginary eigenfrequency of the unstable fixed point. For this Hamiltonian and for the unstable 
point $\theta=0$, 
\begin{equation}
\label{eq:lambda_ent}
\lambda_{h}=2\sqrt{h(J-h)} \ .
\end{equation}
At finite size $N$, the entanglement entropy is bounded and thus always saturates to a finite value, as in Eq.~(\ref{eq:sat}). 
For $N_A=N/2$ this corresponds to $\log \sqrt N$, as shown in the inset of Fig.~\ref{fig:3}(left). 
Conversely, in Fig.~\ref{fig:3}(right), we plot the entanglement entropy dynamics for various possible bipartitions, i.e., various fractions $f_A$ of spins in subsystem $A$, and we compare it with the numerically exact results at fixed $N$. The latter reproduces the former up to $t_{ \text{Ehr}}$, when it saturates around the predicted value $1/2\log N_A$.   
\begin{figure}[t]
\fontsize{12}{10}\selectfont
\centering
\includegraphics[width= 0.45 \textwidth]{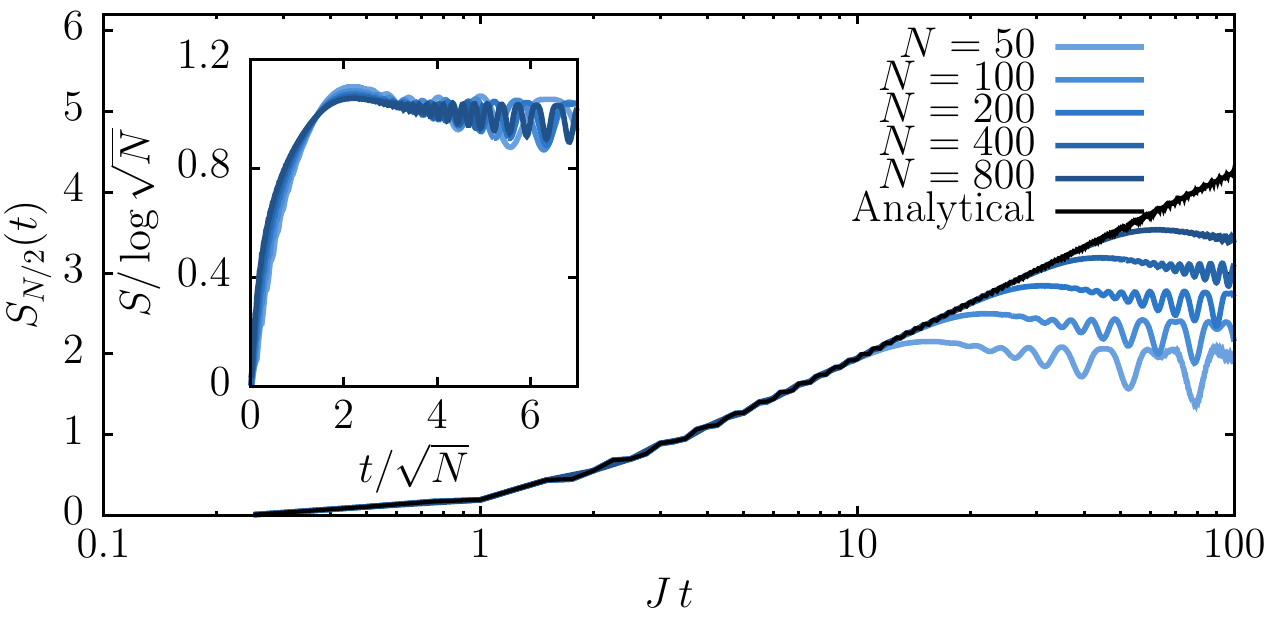}
\includegraphics[width= 0.45 \textwidth]{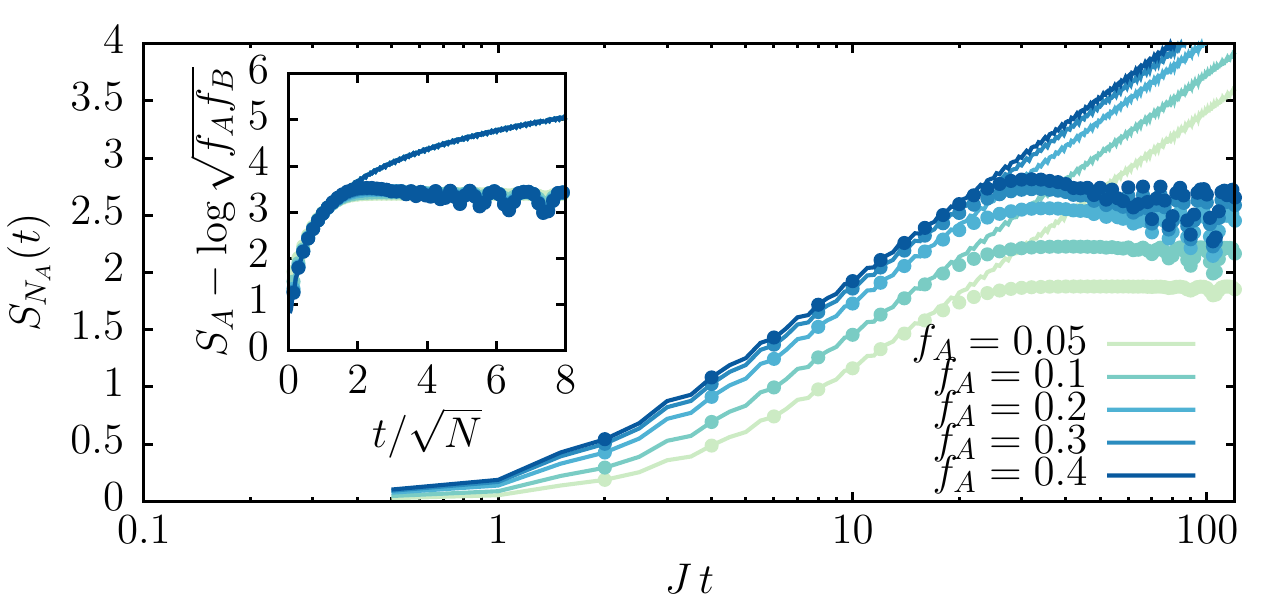}
\caption{Entanglement entropy dynamics after a quench dynamics from $h_0=0$ to $h_f=2J$. (Left)  We compare our analytic formula (black line) for the half-system entanglement entropy $S_{N/2}$ with the ED data for increasing system sizes $N=50\div 800$. The exact diagonalization results follow the logarithmic growth 
up to $t_{ \text{Ehr}}\sim \sqrt N$, where they saturate to $S_{N/2}\sim \log N$. The inset shows the same data with $S_{N/2}$ rescaled by $\log N$ and time by $\sqrt N$. (Right) $S_{N_A}(t)$ for various bipartitions with fractions of spins $f_A= N_A/N =0.05\div0.4$ and fixed size $N=200$. Analytical results from  Eq.~(\ref{eq:EE}) (full lines) are compared with exact numerical results (dots). 
In the inset, $S_{N_A}-1/2\log f_A f_B$ is plotted as a function of the rescaled time $t/\sqrt N$, in order to highlight the validity of the expansion in Eq.~(\ref{eq:big_S}). }
\label{fig:3}
\end{figure}

\begin{figure}[t]
\fontsize{12}{10}\selectfont
\centering
\includegraphics[width= 0.27 \textwidth]{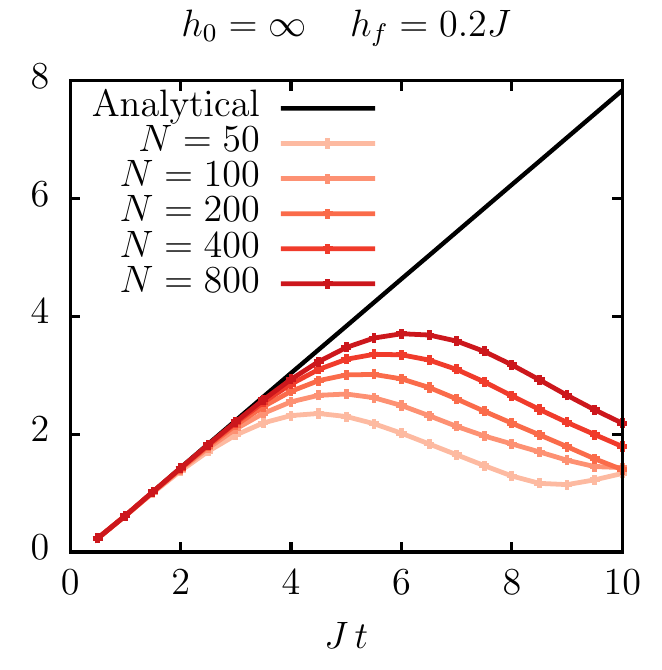}
\includegraphics[width= 0.267 \textwidth]{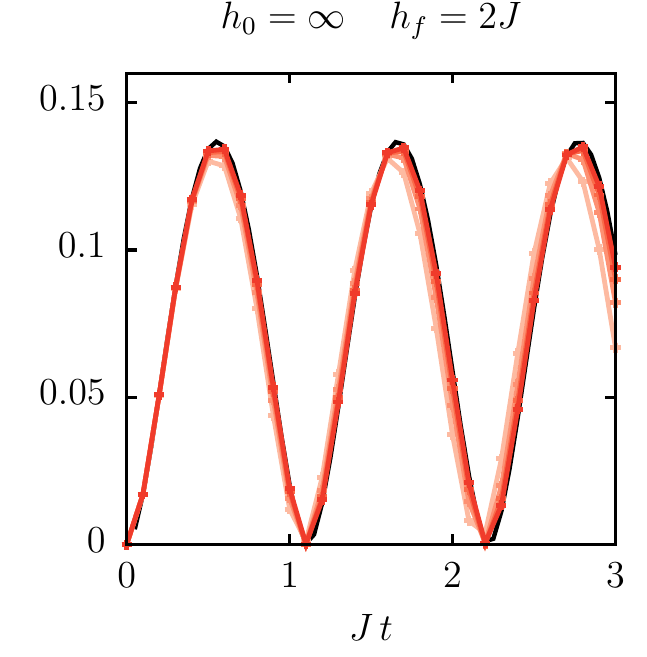}
\caption{Half-system entanglement entropy dynamics after a quench from the paramagnetic ground state $\ket{\phi_0}$. Analytical results from  Eq.~(\ref{eq:EE}) (black lines) are compared with exact numerical results (colours) at finite $N=50\div 800$.  (Left) $h_f=0.2 J$ linear growth of $S_A\sim \lambda_{h_f} t$ with $\lambda_{h_f}$ given in Eq. (\ref{eq:lambda_ent}). (Right) $h_f=2J$ periodic oscillations of $S_A(t)$ resulting from the periodic dynamics of the quantum collective fluctuations.
}
\label{fig:2SM}
\end{figure}

Let us now discuss the opposite case of a quantum quench from a paramagnetic initial state, i.e., 
from $h_0=\infty$
\beq \ket{\phi_0}=\ket{\uparrow \, \uparrow \, \dots \uparrow \,},\eeq 
which corresponds to the initial conditions $\theta_0=0$, $G^{QP}(t=0)=0$ and $G^{QQ}(t=0)=G^{PP}(t=0)=1/2$  in Eqs.~(\ref{eq:motion_angles0}), (\ref{eq:motion_feedback_0})  (due to the singularity of spherical coordinates, the value of $\phi_0$ is immaterial in this case). 
For $h_f<J$, the initial state lies on top of the unstable trajectory at $\theta=0$, hence the collective quantum fluctuations grow exponentially in time. The theory then predicts a linear increase $S_A\sim \lambda_{h_f} t$ of entanglement before $t_{\text{Ehr}}$, with $\lambda_{h_f}$ given in Eq. (\ref{eq:lambda_ent}), see Fig.~\ref{fig:2SM} (left). Conversely, for $h_f>J$, the quantum fluctuations of the initial state undergo oscillations with period $T_{\text{cl}}$, which leads to a periodic dynamics of entanglement entropy as in Fig.~\ref{fig:2SM} (right), with a simple semiclassical interpretation in terms of periodic squeezing of the collective spin. \\

\paragraph{Numerical results for $0<\alpha\leq 1$.}

\begin{figure}[t]
\fontsize{12}{10}\selectfont
\centering
\includegraphics[width= 0.4 \textwidth]{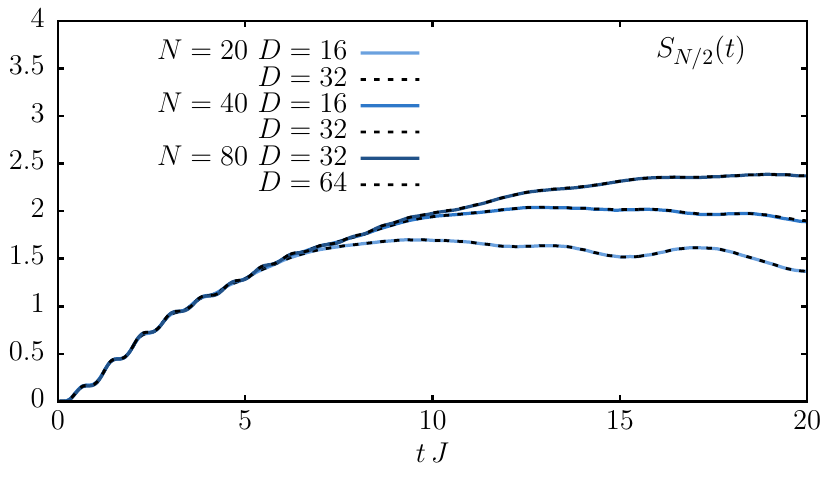}
\includegraphics[width= 0.4 \textwidth]{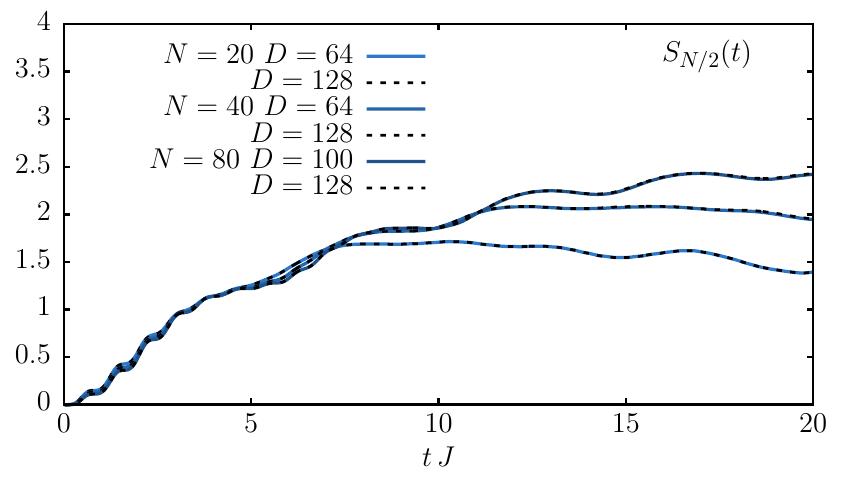}
\caption{Convergence of the MPS-TDVP data upon increasing the bond dimension $D$, for $\alpha=0.1$ (left panel) and $0.7$ (right panel) for the quench $h_0=0 \to h_f=2J$. For each system size $N=20\div 80$, full and dashed lines represent two values of $D$.
}
\label{fig:3SM}
\end{figure}

For spatially-decaying interactions, we simulate the entanglement dynamics adopting the matrix-product-state time-dependent variational principle (MPS-TDVP) \cite{Haegeman2016unif, Haegeman2011timedepe, frownies2010tensor}, a well-suited technique for the study of long-range interacting spin systems \cite{HaukeTagliacozzo, buyskikh2016entanglement, Zunkovic2016}. We approximate the long-range matrix-product-operator (MPO) with a sum of exponentials, as detailed in Ref. \onlinecite{pirvu2010matrix}, with a relative tolerance of $10^{-11}$. Then, for fixed system size, we increase the MPS bond dimension $D$ up to convergence in the desired time window, as shown in Fig.~\ref{fig:3SM}.

In the main text, the converged results are compared with the $\alpha$-independent collective spin-squeezing contribution in Eq.~(\ref{eq:EE}) {and with the full ($\alpha$-dependent) spin-wave entanglement entropy converged with respect to increasing~$N$. 
In order to compute the latter, 
we derive from the Hamiltonian (\ref{eq:Hsw}) the following evolution equations for the spin-wave correlations in $k$-space, defined by $\tilde{G}_{{k}}^{\alpha\beta}(t) = 
\Big\langle  
\tilde{\alpha}_{k} (t) \tilde{\beta}_{-{k}} (t) + \tilde{\beta}_{{k}} (t) \tilde{\alpha}_{-{k}} (t)
\Big\rangle/2$ for $\alpha,\beta=q,p$:
\begin{equation}
\label{eq:motion_feedback}
\left\{
\begin{split}
\dot{G}_{{k}}^{qq}  = \, & 4 \tilde J_k\,   \cos\theta(t) \cos\phi(t)\sin\phi(t) \, \, \tilde{G}_{{k}}^{qq}
+4\left(J \cos^2\phi(t) -\tilde J_k\,  \sin^2\phi(t) \right)\, \tilde{G}_{{k}}^{qp},\\
\dot {G}_{{k}}^{pp} = & -4\left( J \cos^2\phi(t) - \tilde J_k\,\, \cos^2\theta(t) \cos^2\phi(t) \right) \tilde{G}_{{k}}^{qp} 
-4 \tilde J_k\, \,  \cos\theta(t) \cos\phi(t)\sin\phi(t) \,  \tilde{G}_{{k}}^{pp},\\
\dot {G}_{{k}}^{pq} = & -2\left(J\cos^2\phi(t) - \tilde J_k\, \,   \cos^2\theta(t) \cos^2\phi(t) \right)\tilde{G}_{{k}}^{qq}
+ 2 \left(J \cos^2\phi(t) -\tilde J_k\,\, \sin^2\phi(t) \right) \tilde{G}_{{k}}^{pp},
\end{split} 
\right.
\end{equation}
where  $\tilde J_k \equiv J \widetilde{f}_{\alpha,k}$ (see Fig. \ref{fig_ftilde}) and the periodic time-dependence of the angles $\theta(t)$, $\phi(t)$ is determined by the classical precession of the collective spin described by Eq. \eqref{eq:motion_angles0}.
The Fourier antitransform of $\tilde{G}_{{k}}^{\alpha\beta}(t)$ gives the real space correlation matrix in Eq. (\ref{eq:corre_matrix}), whence one computes the time-dependent entanglement entropy using Eq. (\ref{eq:ee_sw}).}

\begin{figure}[H]
\fontsize{12}{10}\selectfont
\centering
\includegraphics[scale = 1]{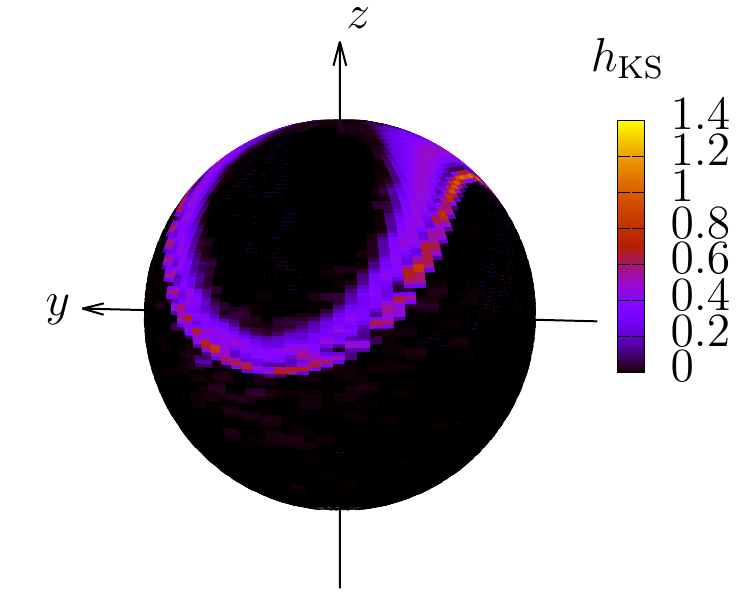}
\caption{
{Density plot of the Kolmogorov-Sinai entropy rate $h_{KS}(\theta_0, \phi_0)$ for different initial conditions $(\theta_0, \phi_0)$ on the Bloch sphere for $\alpha=0.7$, $h=0.5J$. The picture is converged with respect to refining the $k$-space discretization.} 
 }
\label{fig:sferettaIsing}
\end{figure}

{
Following the discussion in Sec. \ref{sec_decaying}C, we first perform a global stability analysis of the spin-wave dynamics 
upon varying the range of interactions, the Hamiltonian parameters and the {initial configuration}. 
For illustration purposes, we focus on quenches to the ferromagnetic phase with $\alpha=0.7$ and $h=0.5 J$ starting from generic fully-polarized spin states, characterized by angles $(\theta_0,\phi_0)$ spanning the entire Bloch sphere.
The spherical plot in Fig. \ref{fig:sferettaIsing}, shows the value of $h_{KS}(\theta_0, \phi_0) = \sum_{k} \big\lvert \Re [\lambda_k(\theta_0, \phi_0)]  \big\rvert$  as a function of the initial configuration on the Bloch sphere.
This quantity is vanishing for regular evolutions free of dynamically unstable excitation modes (i.e., with an entirely real Floquet dispersion relation $ \{ - i \lambda_k \equiv \omega_k \}_k$), whereas it is strictly positive  when at least one mode $k^*$ is resonantly excited [i.e., $\Re (\lambda_{k^*} ) \neq 0$].
The results show that only quenches near dynamical criticality give rise to resonant excitation of spin waves --- i.e., initial configurations close to the classical separatrix of the mean-field  dynamics, see the purple region on the sphere in Fig. \ref{fig:sferettaIsing}, left panel. 
In other words, the single unstable trajectory for $\alpha=0$ broadens to a finite layer of instability for $\alpha>0$.
However, for typical quenches well away from criticality (black region on the sphere in Fig. \ref{fig:sferettaIsing}, left panel), spin waves are non-resonantly excited.
}

\begin{figure}[H]
\fontsize{12}{10}\selectfont
\centering
\includegraphics[width= 0.4 \textwidth]{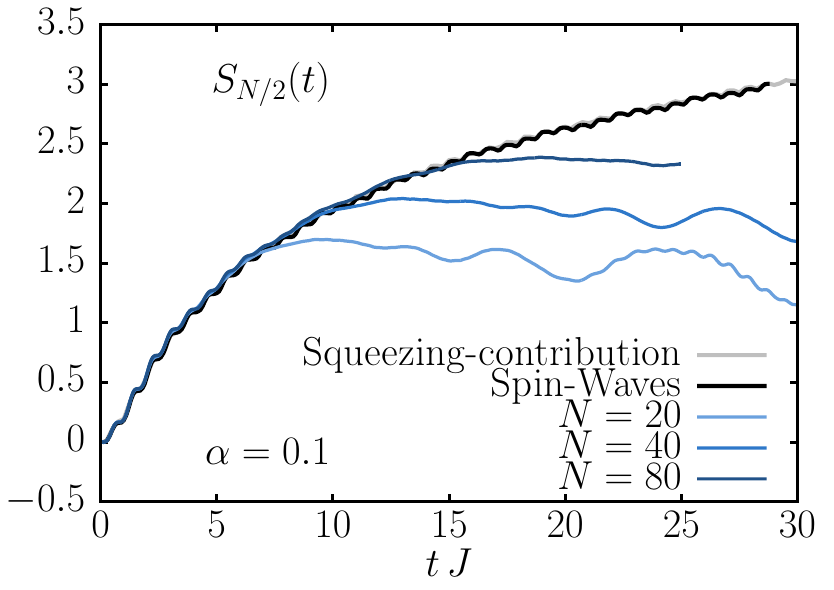}
\includegraphics[width= 0.4 \textwidth]{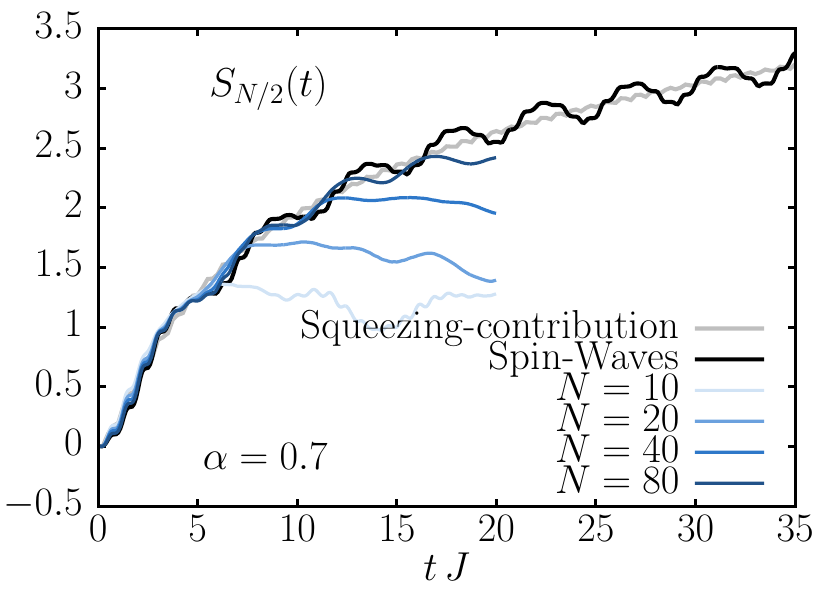}
\caption{
{Comparison between finite-size MPS-TDVP numerical data (light-to-dark blue curves for increasing $N$), the spin-squeezing contribution (grey) and full spin-wave entanglement (black), for $\alpha=0.1$ (left panel) and $0.7$ (right panel), for the quench $h_0=0 \to h_f=2J$.
Numerical data exhibit convergence to the spin-wave result as $N\to\infty$, increasingly more slowly as $\alpha$ is raised from $0$ to $1$ (cf. Fig. \ref{fig_ftilde}, bottom right panel).}
}
\label{fig:4SM}
\end{figure}
\begin{figure}[H]
\fontsize{12}{10}\selectfont
\centering
\includegraphics[width= 0.4 \textwidth]{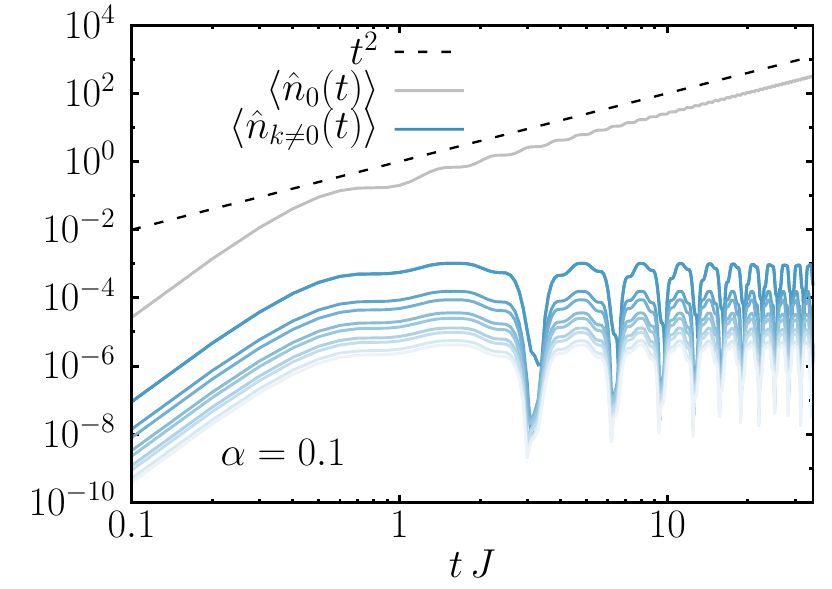}
\includegraphics[width= 0.4 \textwidth]{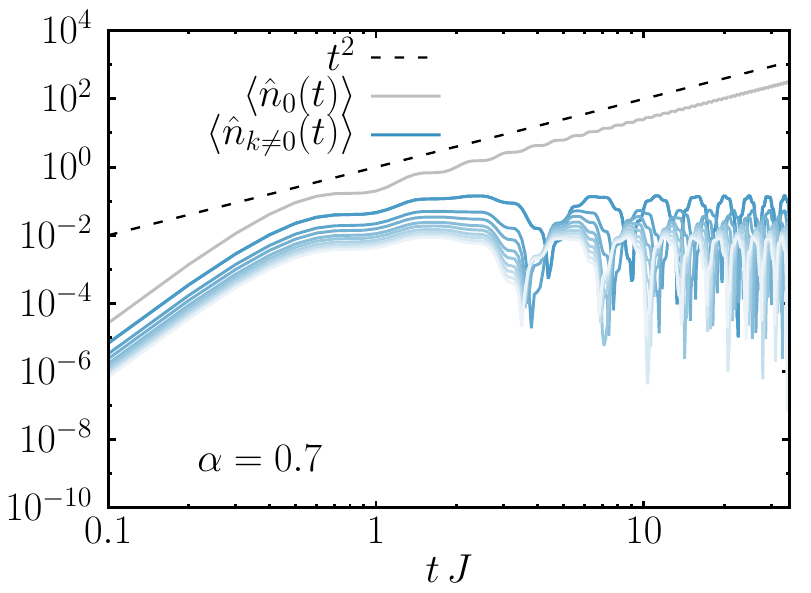}
\caption{
{Time-dependent $k$-resolved spin-wave population for $\alpha=0.1$ (left panel) and $\alpha=0.7$ (right panel) after a quench from $h_0=0$ to $h_f=2J$. The blue color gradient for the spin-wave populations in Fourier modes follows the quasimomentum $\abs{k}$ from the darkest ($k=\pm 2\pi/L$) down to smaller-wavelength modes with larger $\abs{k}$  (only the first $20$ modes are shown). }
}
\label{fig:populationIsing}
\end{figure}

{
As results from the discussion in Sec. \ref{sec_decaying}C and from extending the above analysis to varying $\alpha$ and $h$, we find that
for typical quenches the population $\langle \hat n_{k\neq 0} \rangle$ of spin waves 
remains bounded in time, whereas the collective fluctuations $\langle \hat n_{k=0} \rangle \equiv \langle \hat n_{\text{exc}} \rangle$ grow polynomially in time as $t^2$, as proved in Sec. \ref{sec:neqgrowth}.
Consequently, 
entanglement entropy growth is dominated by the spin-squeezing contribution, and spin-wave excitations generate a finite correction, the impact of which grows as $\alpha$ increases.
This scenario is illustrated in Fig. 3, bottom panels, of the Letter, shown also in Fig. \ref{fig:4SM} here on a linear time scale.
These plots highlight that: 
\textit{i}) the spin-squeezing contribution captures the leading behavior of the time-dependent entanglement entropy for all $\alpha<1$, provided long-wavelength modes are non-resonantly excited, as occurs in typical quenches;  
\textit{ii}) the full spin-wave entanglement entropy calculation \emph{quantitatively} reproduces the numerical data obtained via the MPS-TDVP with periodic boundary conditions,  before saturation as $N$ is increased.
These two occurrences strongly support the effectiveness of the nonequilibrium spin-wave analysis of entanglement entropy growth.
In order to fully corroborate the above picture, we report in Fig. \ref{fig:populationIsing}  the time evolution of the $k$-resolved spin wave population  for the same quench.  }

{Finally, we have explicitly checked that the observation  in Ref. \onlinecite{mori2018prethermalization} of exponentially growing long-wavelength correlations and of semiclassical chaos  for $\alpha > 0$ made therein, is in full agreement with the theory developed here, as the considered initial configuration  falls inside a layer of instability near the mean-field dynamical critical trajectory of the considered Hamiltonian (analogous to the purple region in Fig. \ref{fig:sferettaIsing}). In particular, entanglement is predicted (and verified) to grow linearly in time for these quenches, whereas for most  spin-polarized initial configurations evolving with the same Hamiltonian, the growth remains logarithmic in time.
}


\clearpage 
\bibliography{biblio2}  